\documentclass[epj]{svjour}
\usepackage{graphics}
\usepackage{url}
\usepackage{color}
\usepackage{amsmath}

\def\es0{$E_{\rm sym}(\rho_0)$}

\def\us0{$U_{\rm sym}(\rho_0,k_F)$~}

\def\l0{$L(\rho_0)$~}

\begin{document}
\title{Impact of the nuclear equation of state on the formation of twin stars}
\author{Nai-Bo Zhang\inst{1} and Bao-An Li\inst{2}
}                     
\mail{naibozhang@seu.edu.cn; Bao-An.Li@Tamuc.edu}          
\institute{School of Physics, Southeast University, Nanjing 211189, China \and Department of Physics and Astronomy, East Texas A$\&$M University, Commerce, TX 75429, USA}
\date{Received: date / Revised version: date}
%
\abstract{Twin stars-two stable neutron stars (NSs) with the same mass but different radii have long been proposed to appear as a consequence of a possible first-order phase transition in NS matter. Within a meta-model for the EOS of hybrid stars,  we revisit the viability of twin stars and its dependence on numerous parameters characterizing the EOS of nuclear matter, quark matter, and the phase transition between them. While essentially no experimental constraint exists for the last two, parameters characterizing the EOS of neutron-rich nucleonic matter have been constrained within various ranges by terrestrial experiments and astrophysical observations. Within these ranges, the impact of nuclear EOS and crust-core transition density on the formation of twin stars is studied. It is found that the symmetry energy of neutron-rich nucleonic matter notably influences the formation of twin stars, particularly through its slope $L$ and curvature $K_{\rm sym}$. Conversely, varying the EOS of symmetric nuclear matter within their currently known uncertainty ranges shows minimal influence on the formation of twin stars.
}
\authorrunning{Zhang \& Li}
\titlerunning{Sound Speeds and Spinodal Decompositions in Dense Neutron-Rich Matter}
\maketitle
\section{Introduction}
Neutron star (NS) cores contain the densest visible matter in the universe. At such extreme densities, the NS matter containing only hadrons and leptons may become unstable, undergoing a phase transition to quark matter and thus enabling the formation of hybrid stars. However, to understand properties of supradense matter especially its phase transition, such as its critical hadron-quark transition baryon density $\rho_t$, energy density discontinuity ${\rm \Delta} \varepsilon$, and the stiffness of quark matter measured in terms of its speed of sound squared $c_{\rm s}^2$, remains an outstanding challenge in nuclear astrophysics. In particular, four possible topologies of the mass-radius relation for hybrid stars were suggested in Ref. \cite{Alford13}. Their properties have been investigated using various EOS models in the literature. Nevertheless, many interesting issues regarding the impact of nuclear EOS on the formation of twin stars and their properties call for further investigations.

\begin{figure}[ht]
  \centering
   \resizebox{0.4\textwidth}{!}{
  \includegraphics{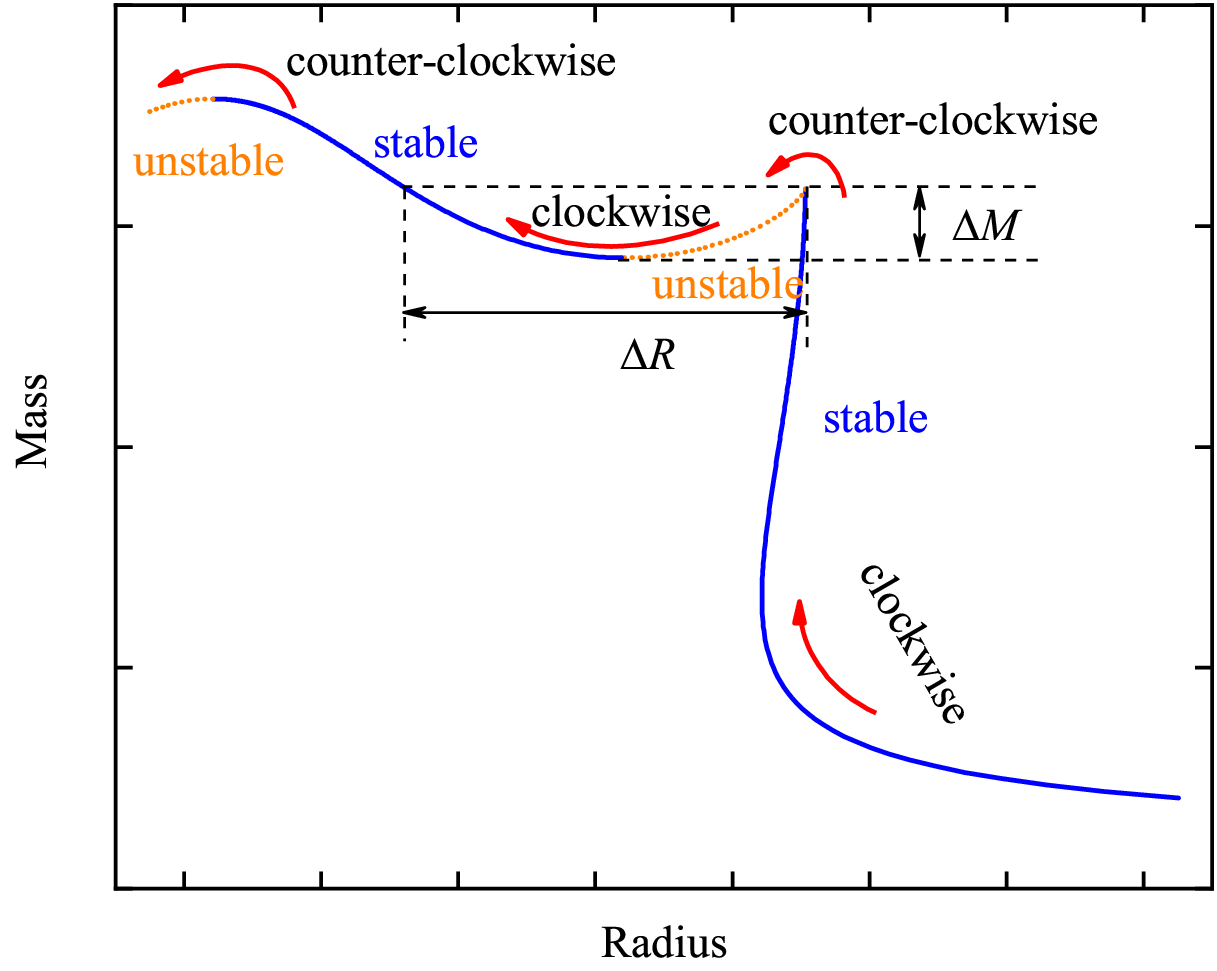}
  }
  \caption{A typical mass-radius relation for twin stars. ${\rm \Delta} M$ quantifies the mass range to form twin stars while ${\rm \Delta} R$ indicates their maximum radius separation.}\label{sketch}
\end{figure}

A possible mass-radius relation of hybrid stars is shown in Fig. \ref{sketch} for introducing some terminologies and quantities characterizing twin stars. During a strong first-order phase transition from nuclear to quark matter, the sudden softening of the EOS results in the appearance of an extremum mass (the first extremum). As pressure increases, the mass initially decreases until reaching a minimum (the second extremum), after which it rises to a second maximum (the third extremum) mass. This non-monotonic mass-radius relationship, caused by the phase transition, can lead to the instability of NSs. The stability of solutions to the Tolman-Oppenheimer-Volkoff (TOV) equation can be assessed using the Bardeen, Thorne, and Meltzer stability criteria \cite{Bardeen66,Harrison65}: At each extremum with increasing central pressure, one stable (unstable) mode becomes unstable (stable) when the mass-radius curve rotates counter-clockwise (clockwise). Within the mass range ${\rm \Delta} M$ between the first two extremums as shown in Fig. \ref{sketch}, two stable NSs with the same mass but different radii may coexist, forming twin stars \cite{Gerlach68,Kampfer81,Kampfer81b,Glendenning20,Schertler20}. The twin stars have a maximum radius separation ${\rm \Delta} R$ at the maximum mass as indicated in Fig. \ref{sketch}. While at the lower boundary of ${\rm \Delta} M$, twin stars have the minimum radius separation. Clearly, the ${\rm \Delta} M$ quantifies the mass range to form twin stars while the ${\rm \Delta} R$ indicates their maximum radius separation. To identify confidently twin stars, ${\rm \Delta} R$ has to be larger than about $2\sigma_{\rm{obs}}$ that is presently about 2 to 4 km based on NICER's observations of PSR J0030+0451 and PSR J0740+6620  \cite{Miller19,Riley19,Miller21,Riley21} as well as LIGO/VIRGO's observation of GW170817 \cite{LIGO18}.
Thus, comparing ${\rm \Delta} R$ obtained with various EOSs with the current precision $\sigma_{\rm{obs}}$ of NS radius measurement may reveal conservatively the difficulties or likelihood of observing twin stars. Moreover, observing twin stars could help confirm the existence of a phase transition in supradense NS matter.

Recent astrophysical observations have stimulated more extensive explorations of twin star properties \cite{Alford16,Ranea16,Montana19,Espino22,Christian20,Christian21,Christian22,Tsaloukidis23,LiJJ21,LiJJ23,LiJJ24,Gorda23,Sun23,Pradhan23,Imajo24,Carlomagno24}. However, their existence remains inconclusive. For example, in our previous studies in either Bayesian inference \cite{Xie21} or forward-modeling \cite{Zhang23} using a meta-model EOS for nuclear matter and the constant speed of sound (CSS) model for quark matter we found that twin stars cannot satisfy the observational constraints from NICER \cite{Riley19,Miller19,Riley21,Miller21} and GW170817 \cite{LIGO18} simultaneously. In particular, in the forward-modeling approach \cite{Zhang23} when all nuclear matter parameters are fixed at their presently known most probable values (MPVs) based on previous analyses of many terrestrial nuclear experiments and recent NS observations, no evidence for the formation of twin stars was found. On the other hand, a more recent study \cite{LiJJ24} concluded that low-mass twin stars could satisfy current NS observations. In their work, they varied the slope $L$ of nuclear symmetry energy and the skewness $J_0$ of symmetric nuclear matter (SNM) in relatively large ranges. Particularly, they varied $J_0$ from $-600$ to $1000$ MeV, which is much wider than the constraints of $J_0=-190\pm100$ MeV at 68\% confidence level extracted from observations of NS \cite{Zhang19,Xie19,Xie20} and $J_0=-180^{+100}_{-110}$ MeV from analyzing nuclear collective flow in intermediate-high energy heavy-ion reactions \cite{Xie21JPG}. Thus, it would be interesting to revisit the issue of forming possibly twin stars within the same meta-model EOSs as in our previous work \cite{Xie21,Zhang23} but within the currently allowed whole space of EOS parameters instead of with their most probable values only.
Indeed, we found that twin stars can exist in some EOS parameter space. Quantitatively, we examine the ${\rm \Delta} M$
individually as a function of several key EOS parameters in their currently known uncertainty ranges.

The rest of this paper is organized as follows: The meta-model EOSs describing nuclear matter and quark matter as well as the transition between them are introduced in Sec. II. Effects of the EOS parameters and crust-core transition density on the formation of twin stars are discussed in detail in Sec. III. Our findings are summarized in Sec. IV.

\section{Construction of equations of state for hybrid stars}

In the present work, the EOS of nuclear matter ($npe\mu$) in NSs at $\beta$-equilibrium is described by a meta-model
that is a model of models, see, e.g. Ref. \cite{Wang}. An NS meta-model provides a framework constructed by using EOS parameters that can be sampled randomly within their currently known broad uncertain ranges to mimic essentially all existing NS EOS models in the literature. The EOS of quark matter and its connection with nuclear matter in NSs is described by the CSS model. The latter is also a meta-model (in the sense that it is a model of models by varying its parameters). The two phases of NS matter are connected through a first-order phase transition with an energy density gap ${\rm \Delta}\varepsilon$ and a transition density $\rho_t$. For the sake of completeness and clarity in subsequent discussions, we briefly recall the main features of these models. More details can be found in the literature as we shall point out in more detail.

\subsection{An EOS for nuclear matter in neutron stars}
In Ref. \cite{Zhang18}, we constructed a parameterized EOS of $npe\mu$ matter at $\beta$-equilibrium in NSs by parameterizing separately the EOS of SNM $E_{0}(\rho)$ and nuclear symmetry energy $E_{\rm sym}(\rho)$ as:
\begin{equation}\label{E0-taylor}
  E_{0}(\rho)=E_0(\rho_0)+\frac{K_0}{2}(\frac{\rho-\rho_0}{3\rho_0})^2+\frac{J_0}{6}(\frac{\rho-\rho_0}{3\rho_0})^3,
\end{equation}
\begin{eqnarray}\label{Esym-taylor}
    E_{\rm{sym}}(\rho)&=&E_{\rm{sym}}(\rho_0)+L(\frac{\rho-\rho_0}{3\rho_0})\nonumber\\
    &+&\frac{K_{\rm{sym}}}{2}(\frac{\rho-\rho_0}{3\rho_0})^2
  +\frac{J_{\rm{sym}}}{6}(\frac{\rho-\rho_0}{3\rho_0})^3.\nonumber\\
\end{eqnarray}
More details about this model can be found in our previous publications \cite{Zhang19a,Zhang19b,Zhang2020,Zhang2021,Zhang22,Xie24}.

The binding energy $E_0(\rho_0)$ and incompressibility $K_0$ of SNM at the saturation density have been relatively well constrained to $E_0(\rho_0)=-15.9\pm0.4$ MeV and $K_0=240\pm20$ MeV \cite{Garg18,Shlomo06}, while the magnitude $E_{\rm sym}(\rho_0)$ and slope $L$ of symmetry energy at $\rho_0$ are constrained to $E_{\rm sym}(\rho_0)=31.7\pm3.2$ MeV and $L=58.7\pm28.1$ MeV \cite{Li13,Oertel17}, respectively. Regarding parameters characterizing the high-density behavior of nuclear matter, the curvature of the symmetry energy is around $K_{\rm sym}=-100\pm100$ MeV \cite{Li21,Grams22,Margueron18,Mondal17,Somasundaram21}, while the skewness of the SNM EOS is constrained to $J_0=-190\pm100$ MeV \cite{Zhang19,Xie19,Xie20} based on terrestrial experiments and astrophysical observations. Few constraints on $J_{\rm sym}$ have been obtained so far, and it is only very roughly known to be around $-200<J_{\rm sym}<800$ MeV \cite{Cai17,Zhang17,Tews17}.
We emphasize that the above parameters can be varied independently within their uncertain ranges given above. In fact, some of the uncertainty ranges are obtained from the marginalized posterior probability distribution functions (PDFs) of individual parameters in Bayesian analyses of NS observables. Nevertheless, once new observables or physics conditions are considered, correlations among the updated posterior EOS parameters may be introduced. In this work, however, all EOS parameters should be considered independent in their current uncertainty ranges given above.

Once the parameters in Eqs. (\ref{E0-taylor}) and (\ref{Esym-taylor}) are given, the energy density for $npe\mu$ matter (also interchangeably called hadronic matter (HM)) at $\beta$-equilibrium in NSs can be calculated from
\begin{equation}\label{lepton-density}
  \varepsilon_{\rm{HM}}(\rho, \delta)=\rho [E(\rho,\delta)+M_N]+\varepsilon_l(\rho, \delta),
\end{equation}
where $M_N$ represents the average nucleon mass of 938 MeV,
\begin{equation}\label{Erho}
E(\rho,\delta)=E_0(\rho)+E_{\rm{sym}}(\rho)\cdot\delta ^{2}+\mathcal{O}(\delta^4)
\end{equation}
is the EOS of asymmetric nuclear matter (ANM), $\delta=(\rho_n+\rho_p)/\rho$ is the isospin asymmetry, and $\varepsilon_l(\rho, \delta)$ denotes the lepton energy density which can be calculated using the ideal Fermi gas model \cite{Oppenheimer39}. The baryon densities $\rho_i$ of particle $i$ can be obtained by solving the $\beta$-equilibrium condition $\mu_n-\mu_p=\mu_e=\mu_\mu\approx4\delta E_{\rm{sym}}(\rho)$ where $\mu_i=\partial\varepsilon(\rho,\delta)/\partial\rho_i$ and charge neutrality condition $\rho_p=\rho_e+\rho_\mu$. Then the energy density $\varepsilon$ and pressure $P$ both become barotropic, i.e. a function of density only. In particular, the pressure as a function of density only can be calculated from:
\begin{equation}\label{pressure}
P(\rho)=\rho^2\frac{{\rm d}\varepsilon(\rho,\delta(\rho))/\rho}{{\rm d}\rho}.
\end{equation}
Based on the above equations, we can then obtain a unique EOS in the form of $P(\varepsilon)$ for the hadronic phase of NS matter once the parameters are fixed.

We adopt the Negele-Vautherin (NV) EOS \cite{Negele73} for the inner crust and the Baym-Pethick-Sutherland (BPS) EOS \cite{Baym71b} for the outer crust. The crust-core transition density, $\rho_c$, can either be fixed at a fiducial value ($\rho_c=0.08$ fm$^{-3}$) or determined through alternative methods. Here we use a thermodynamic approach \cite{Kubis04,Kubis07,Lattimer07}, where $\rho_c$ is obtained by identifying the density at which the incompressibility of neutron star matter in the uniform core vanishes:
\begin{eqnarray}\label{Kmu}
    K_\mu&=&\rho^2\frac{{\rm d^2}E_0}{{\rm d}\rho^2}+2\rho\frac{{\rm d}E_0}{{\rm d}\rho}+\delta^2\nonumber\\
    &\times&[\rho^2\frac{{\rm d^2}E_{\rm sym}}{{\rm d}\rho^2}+2\rho\frac{{\rm d}E_{\rm sym}}{{\rm d}\rho}-2E_{\rm sym}^{-1}(\rho\frac{{\rm d}E_0}{{\rm d}\rho})^2]=0.\nonumber\\
\end{eqnarray}
Since $\rho_c$ can influence the radius of a neutron star, its effects on the formation of twin stars should be carefully examined.
For this purposes, we shall compare results obtained with $\rho_c=0.08$ fm$^{-3}$ and the one from the thermodynamic approach described above. 

\subsection{A meta-model for nuclear-quark phase transition and quark matter}

To describe the possible quark matter in the cores of hybrid stars, we adopt the CSS model of Alford, Han, and Prakash assuming the phase transition from nuclear matter to quark matter is first-order \cite{Alford13,Chamel13,Zdunik13}. In this model, the phase transition and the EOS of quark matter can be described by:
\begin{equation}
\varepsilon(\rho)= \begin{cases}\varepsilon_{\mathrm{HM}}(\rho), & \rho<\rho_{t} \\ \varepsilon_{\mathrm{HM}}\left(\rho_{t}\right)+{\rm \Delta} \varepsilon+c_{\mathrm{s}}^{-2}\left(p-p_{t}\right), & \rho>\rho_{t}\end{cases}
\end{equation}
where $\varepsilon_{\mathrm{HM}}(p)$ is the energy density of hadronic matter (HM) described in the previous subsection, $p_t$ is the pressure at the transition density $\rho_t$, while ${\rm \Delta} \varepsilon$ is the gap in energy density between the hadronic and quark phases. It has been shown in the literature that the CSS model can capture the characteristics of various microscopic quark matter models, such as Nambu-Jona-Lasinio (NJL) models \cite{Zdunik13,Agrawal10,Bonanno12,Lastowiecki12}, perturbation theories \cite{Kurkela10,Kurkela10b}, or bag-model-like EOSs \cite{Traversi21,Traversi20}. It has been extensively employed in studying hybrid stars \cite{Alford13,LiJJ21,Xie21,Alford15,Ayriyan15,Drischler22,Chatziioannou20,Han20,Li22,Miao20}.

Note that the CSS model is by no means the only model describing quark matter that can satisfy all the presently available astrophysical constraints, while it provides a physically robust framework that is technically very flexible in describing quark matter in the cores of hybrid stars. Several alternative models have been developed to describe the properties of quark matter, each with distinct assumptions and features. For instance, the NJL model \cite{Zdunik13,Agrawal10,Bonanno12,Lastowiecki12}, the density-dependent quark model \cite{Backes21,Wen05}, and the mean field theory of quantum chromodynamics (QCD) EOS model \cite{Albino24,Roy24} can all simulate hybrid stars that satisfy observational constraints from NICER \cite{Miller19,Riley19,Miller21,Riley21} and GW170817 \cite{LIGO18}. Despite their differences, these models share the common goal of providing a realistic EOS for quark matter, which is crucial for accurately describing the internal structure and observational properties of hybrid stars. The CSS model is particularly attractive for its versatility, as it can encompass features of several of these microscopic models while remaining analytically tractable.

Once the EOS of nuclear matter is determined, the properties of hybrid stars are solely dictated by the three CSS model parameters. By design, the stiffness of quark matter is controlled by the speed of sound squared $c_s^2$. To support NSs at least as massive as about 2.0 M$_\odot$, we limit $c_s^2$ to the range of $0.5<c_s^2<1$. Similarly, a higher value of ${\rm \Delta} \varepsilon$ leads to a smaller hybrid star mass, while a lower value of ${\rm \Delta} \varepsilon$ complicates the formation of twin stars. We set ${\rm \Delta} \varepsilon$ within the range of $150<{\rm \Delta} \varepsilon<350$ MeV. As for the transition density $\rho_t$, we select $1.5\rho_0<\rho_t<3.5\rho_0$ in this work. It is consistent with the relatively low transition densities found in Refs. \cite{Xie21,Miao20,Liang21,Somasundaram22}. As a summary, we show the most probable values (MPVs) and uncertainties of each parameter mentioned above in Tab. \ref{Tab1}. Unless specifically noted, all parameters are fixed at their MPVs.

\begin{table}[htbp]
\centering
\caption{The most probable values (MPVs) and uncertainties of parameters describing nuclear and quark matter EOSs as well as phase transition properties.}\label{Tab1}
 \begin{tabular}{ccccccc}
  \hline\hline
   Parameters&~~ MPV~&~Lower limit  &~~Upper limit \\
    \hline\\
$J_0$~~(\rm {MeV}) & $-190$ & $-290$ & $-90$ \\
$E_{\mathrm{sym}}(\rho_0)~~(\rm {MeV})$ & 31.7 & 28.5 & 34.9 \\
$L$~~(\rm {MeV}) & 58.7 & 30 & 90 \\
$K_{\mathrm{sym}}~~(\rm {MeV})$  & $-100$ & $-200$ & $0$  \\
$J_{\mathrm{sym}}~~(\rm {MeV})$ & $-$ & -200 & 800 \\
 \hline
$\rho_t/\rho_0$ & $-$ & 1.5 & 3.5 \\
${\rm \Delta} \varepsilon~~(\rm {MeV/fm^{-3}})$ & $-$ & 150 & 350 \\
$c_s^2$ & $-$ & 0.5 & 1 \\
 \hline
 \end{tabular}
\end{table}

Compared to many NS EOSs derived from microscopic and/or phenomenological nuclear many-body theories, as well as piecewise polytropes or spectrum functions frequently used in the literature, we need to emphasize the following aspects of the meta-model EOS introduced above and justifications for using it:

(1) It is well known that the TOV equations are composition blind in the sense that as long as an EOS in the form of pressure vs energy density is given, the mass-radius sequence of neutron stars is uniquely determined regardless of how the EOS is constructed and/or what particles or phases are included. That is why it is enough to start by parameterizing the EOS in piecewise polytropes or spectrum functions that know nothing about the isospin asymmetry or composition/phase of neutron star matter, and their parameters especially for the high-density pieces have no direct connection with the EOS parameters that are determined by terrestrial experiments.  On the other hand, our meta-model EOS based on the nucleon binding energy in neutron-rich matter considers beta-equilibrium and charge neutrality self-consistently necessary to extract information about the high-density nuclear symmetry energy. The latter is particularly important for determining the proton fraction in supradense NS matter, thus the cooling mechanism of protoneutron stars.

(2) In the microscopic and/or phenomenological nuclear many-body theories for NS matter, numerous fundamental physical details are included. However, the parameters within these theories typically exhibit interrelations, making it challenging to isolate the individual effects of each parameter on the properties of NSs. One of the advantages of our meta-model EOS is its ability to independently vary its parameters within their presently known uncertainty ranges as we discussed earlier. Most importantly, the meta-model EOS provides a way to mimic essentially all existing EOSs predicted by basically all nuclear many-body (microscopic and/or phenomenological) theories in the literature by varying the meta-model EOS parameters. The flexibility and diversity provided by such meta-model EOSs (not available or strictly restricted in most of the microscopic and/or phenomenological theories) are essential both technically and physically in solving the inverse structure problems of neutron stars as we have demonstrated in several of our previous publication with both direct inversion by brute force \cite{Zhang19a,Zhang19b,Zhang2020,Zhang2021,Zhang22} and statistical inversion with Bayesian analyses \cite{Xie21,Xie19,Xie20,Xie24}.

(3) Our meta-model EOS has its caveats. The parameterizations given by Eqs. (\ref{E0-taylor}) and (\ref{Esym-taylor}) look like Taylor expansions with the issue of convergence for densities larger than about 1.5$\rho_0$. We emphasize here that our model is different from the Taylor expansions. We demonstrated in great detail how our model is established in Ref. \cite{Zhang19a}. Surely, Taylor expansions become progressively inaccurate for large densities. Therefore, for describing neutron star matter, we parameterize the NS EOS and treat the coefficients in Eqs. (\ref{E0-taylor}) and (\ref{Esym-taylor}) as unknown parameters to be extracted from astrophysical observations or terrestrial experiments, instead of actually expanding some known energy density functionals. As parameterizations, mathematically they can be valid at any density without the convergence issue even at $(2-3) \rho_0$ and beyond.

(4) The EOS model described above assumes that the hadronic part of NS matter consists solely of ($npe\mu$) matter before the hadron-quark phase transition. Consequently, it does not account explicitly for new particles such as hyperons, mesons, and (${\rm \Delta}$)(1232) resonances, leading to the softening (e.g., the hyperon and/or ${\rm \Delta}$ puzzles) of NS matter as suggested in the literature. We emphasize, however, that within our meta-model their effects on twin stars are mimicked by varying the hadron-quark transition density and other EOS parameters, especially the three CSS EOS parameters. Thus, the uncertainties associated with possibly new particles are currently all included in the ranges of the meta-model EOS parameters.

\section{The formation of twin stars}

\begin{figure*}[ht]
  \centering
   \resizebox{0.9\textwidth}{!}{
  \includegraphics{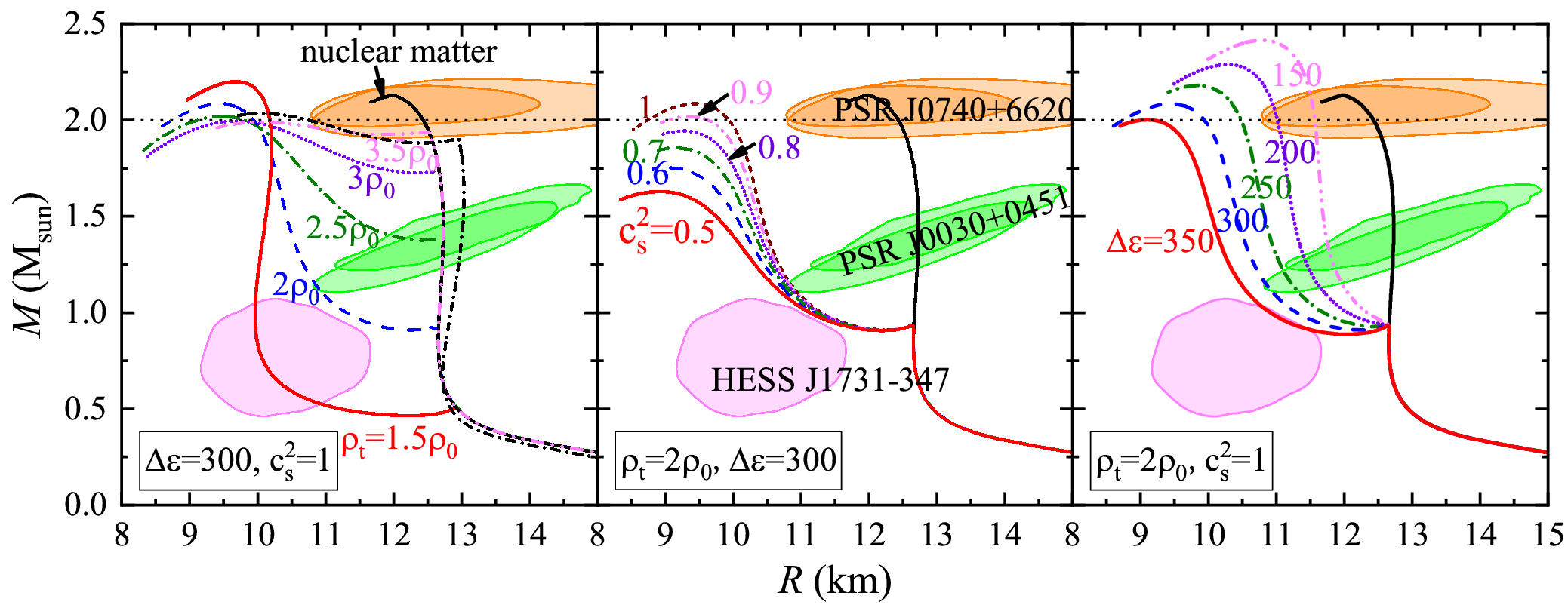}
  }
  \caption{The effects of $\rho_t$ (left panel), $c_s^2$ (middle panel), and ${\rm \Delta}\varepsilon$ (right panel) on the mass-radius relations of a hybrid star. The solid black line represents the mass-radius relation of a traditional NS without a phase transition. The dash-dotted black line in the left panel represents the mass-radius relation of a hybrid star with $c_s^2=1$, ${\rm \Delta}\varepsilon=300$ MeV$\cdot$fm$^{-3}$, $\rho_t=3\rho_0$, $K_{\rm sym}=0$, and $J_0=-100$ MeV. The horizontal dotted line corresponds to $M=2.0$ M$_\odot$. The shaded ranges depict the constraints from NICER on PSR J0030+0451 \cite{Miller19,Riley19} and PSR J0740+6620 \cite{Miller21,Riley21}, as well as the mass-radius relation of HESS J1731-347 \cite{Doroshenko22}.}\label{Figcss}
\end{figure*}

In our previous study \cite{Zhang23}, by fixing the nuclear matter parameters to their most probable values mentioned above, we obtained constraints on the CSS parameters based on astrophysical observables. We found that the twin star is disfavored based on present observations. Going away from the most probable values for the nuclear EOS model parameters, in the following we present results of exploring the possible formation of twin stars in the whole EOS parameter space presently allowed, albeit some areas may have small probabilities to be reached according to some Bayesian analyses we and other have done. Because we have totally 9 EOS parameters, to visualize our results and make the calculations manageable we examine the effects of each parameter individually within its currently known uncertainty ranges while keeping all other parameters at their presently known most probable values. This then limits us to see the effects of those poorly known parameters.

\subsection{The effects of CSS parameters on twin stars}

We first analyze the effects of the CSS parameters describing transition properties on the mass-radius phase diagram of hybrid stars.
In Ref. \cite{Alford13}, the plane of $p_t/\varepsilon_t\sim{\rm \Delta}\varepsilon/\varepsilon_t$ or $\rho_t/\rho_0\sim{\rm \Delta}\varepsilon/\varepsilon_t$ is divided into four ranges. The occurrence of a twin star is possible in the quadrant characterized by relatively small $\rho_t/\rho_0$ and large ${\rm \Delta}\varepsilon/\varepsilon_t$, which varies for different EOSs of nuclear matter (see Ref. \cite{Alford15} for an example). These parameters are chosen to be comparable with the Seidov stability condition \cite{Seidov71,Schaeffer83,Lindblom98} for first-order phase transitions: ${\rm \Delta}\varepsilon/\varepsilon_t\leq1/2+3p_t/2/\varepsilon_t$. The presence of a twin star is generally expected for small $\rho_t$ and large ${\rm \Delta}\varepsilon$ when using the CSS model describing the quark matter \cite{Alford13}.

Shown in Fig. \ref{Figcss} are the effects of varying respectively $\rho_t$ (left panel), $c_s^2$ (middle panel), and ${\rm \Delta}\varepsilon$ (right panel) on the mass-radius diagram of hybrid stars. The solid black line represents the mass-radius relation of a traditional NS without a phase transition when fixing the nuclear matter parameters to their most probable values ($J_{\rm sym}=800$ MeV). The horizontal dotted line corresponds to $M=2.0$ M$_\odot$. The shaded ranges depict the constraints from NICER on PSR J0030+0451 \cite{Miller19,Riley19} and PSR J0740+6620 \cite{Miller21,Riley21}, as well as the mass-radius relation of HESS J1731-347 \cite{Doroshenko22}. It is observed that the traditional NS can successfully satisfy the constraints imposed by NICER. Incorporating the phase transition visibly reduces the radii, making it more challenging to satisfy NICER's constraints. Nevertheless, by adjusting the combinations of CSS parameters and nuclear matter parameters, it is possible to once again satisfy NICER's constraints. As an example, the mass-radius diagram of a hybrid star ($c_s^2=1$, ${\rm \Delta}\varepsilon=300$ MeV$\cdot$fm$^{-3}$, $\rho_t=3\rho_0$, $K_{\rm sym}=0$, and $J_0=-100$ MeV) is shown in the left panel of Fig. \ref{Figcss}. Compared with the dotted purple line of $\rho_t=3\rho_0$, the change of nuclear matter parameters can once again satisfy NICER's constraints. Given that the current study is primarily concerned with parameters influencing the formation of twin stars, we will not enforce the requirement that the theoretical mass-radius relationships should simultaneously describe properly the astrophysical observations in the following discussions.

As shown in the left panel, the condition $M_{\rm TOV}\equiv M_{\rm max}>2$ M$_\odot$ is always satisfied as $\rho_t$ increases from 1.5 $\rho_0$ to 3.5 $\rho_0$, while $M_{\rm TOV}$ correspondingly decreases from 2.2 M$_\odot$ to 2 M$_\odot$. The above relations arise from the stiff EOS of quark matter with $c_s^2=1$, which contributes to the increased maximum mass after a phase transition. The contribution from quark matter becomes more pronounced with decreasing $\rho_t$. As described in the introduction, the mass range in which twin stars can coexist is measured by ${\rm \Delta} M$. Across all $M-R$ curves shown, the appearance of twin stars is evident within a narrow range of ${\rm \Delta} M\leq0.05$ M$_\odot$. As discussed below, the value of $\rho_t$ emerges as a crucial determinant in the formation of twin stars. Additionally, it is clearly shown that twin stars can form at masses around 0.5 M$_\odot$ and 2 M$_\odot$ by varying the transition density. These results indicate that high-mass twin stars are as favored as low-mass ones, which is consistent with previous findings, such as those in Refs. \cite{Typel16,Benic15}, where an excluded volume correction is included in the nuclear EOS at high densities.

The effects of $c_s^2$ on the mass-radius relations are depicted in the middle panel of Fig. \ref{Figcss}. The squared speed of sound of nuclear matter at 2$\rho_0$ is 0.52. Despite the fact that the EOS of the quark matter is stiffer compared to nuclear matter, a significant energy density discontinuity of ${\rm \Delta}\varepsilon=300$ MeV$\cdot$fm$^{-3}$ reduces the maximum mass from 2.09 to 1.63 M$_\odot$ as $c_s^2$ decreases from 1 to 0.5. The effects of $c_s^2$ become noticeable as the mass-radius curves approach the maximum mass, suggesting that $c_s^2$ has a limited impact on the formation of twin stars which normally form near the phase transition point. However, a larger $c_s^2$ would render the formation of twin stars more challenging, as stiffer EOSs with higher $c_s^2$ exhibit slightly greater mass at a fixed radius after a phase transition.

The most important role of CSS parameters in determining the formation of twin stars is displayed in the right panel of Fig. \ref{Figcss}. The variation of ${\rm \Delta}\varepsilon$ from 150 to 350 MeV$\cdot$fm$^{-3}$ not only reduces the maximum mass from 2.41 to 1.99 M$_\odot$ but also gradually introduces the presence of twin stars when ${\rm \Delta}\varepsilon\geq250$ MeV$\cdot$fm$^{-3}$. The large maximum mass of $M_{\rm TOV}=2.41$ M$_\odot$ for ${\rm \Delta}\varepsilon=150$ MeV$\cdot$fm$^{-3}$ arises from the interplay of a stiff EOS of quark matter with $c_s^2=1$ and a small energy density discontinuity ${\rm \Delta}\varepsilon$. Given the apparent impact of ${\rm \Delta}\varepsilon$ on the formation of twin stars, we maintain ${\rm \Delta}\varepsilon=300$ MeV$\cdot$fm$^{-3}$ in the subsequent discussions when demonstrating the effects of nuclear EOS on twin stars.

\subsection{The effects of SNM EOS on twin stars}

\begin{figure}[ht]
  \centering
   \resizebox{0.45\textwidth}{!}{
  \includegraphics{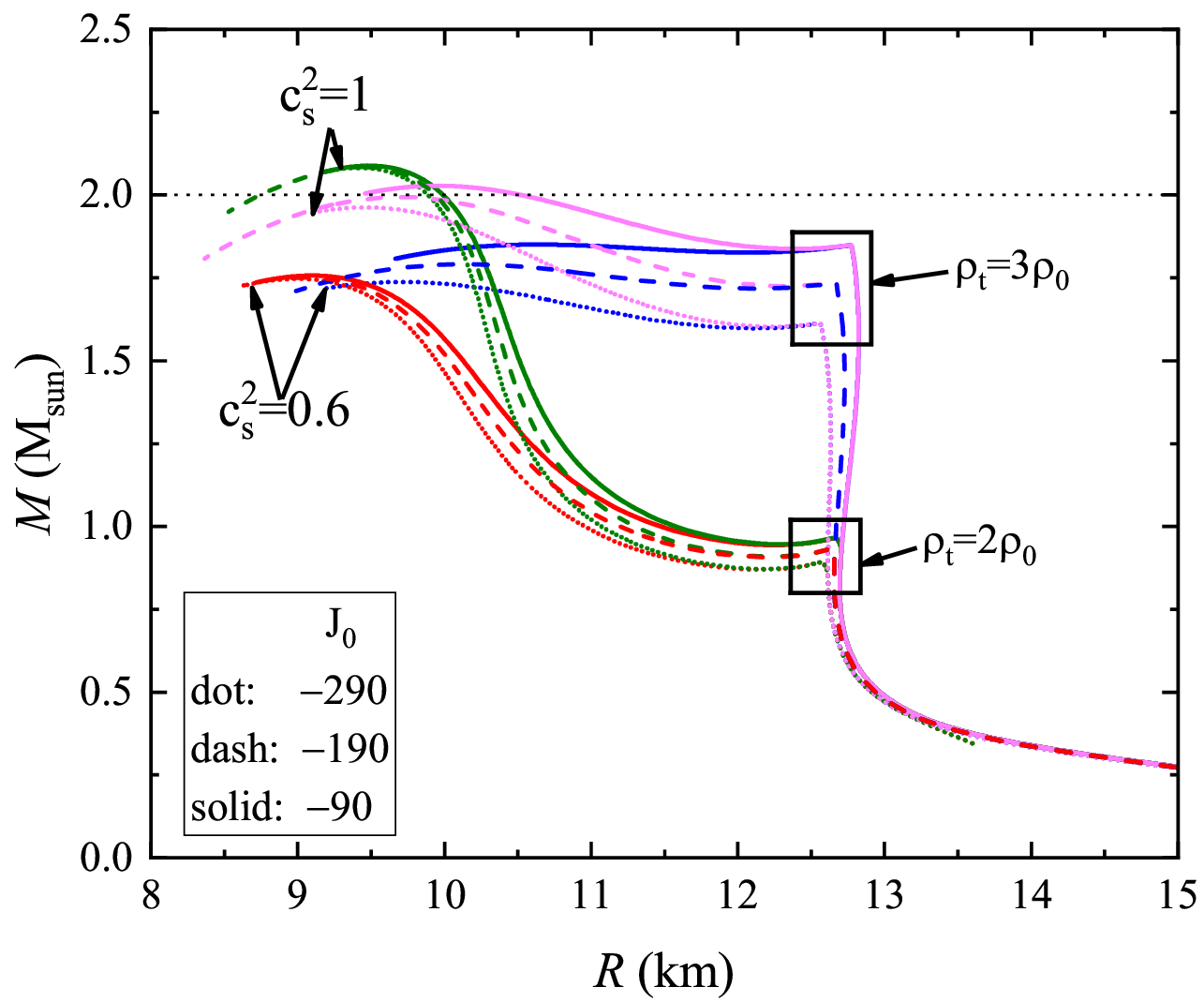}
  }
  \caption{The effects of $J_0$ on the mass-radius relations when $\rho_t=2\rho_0, 3\rho_0$ and $c_s^2=0.6, 1$, respectively. The labeled black boxes mark the ranges where the corresponding phase transition happens.}\label{FigJ0}
\end{figure}

\begin{figure}[ht]
  \centering
   \resizebox{0.4\textwidth}{!}{
  \includegraphics{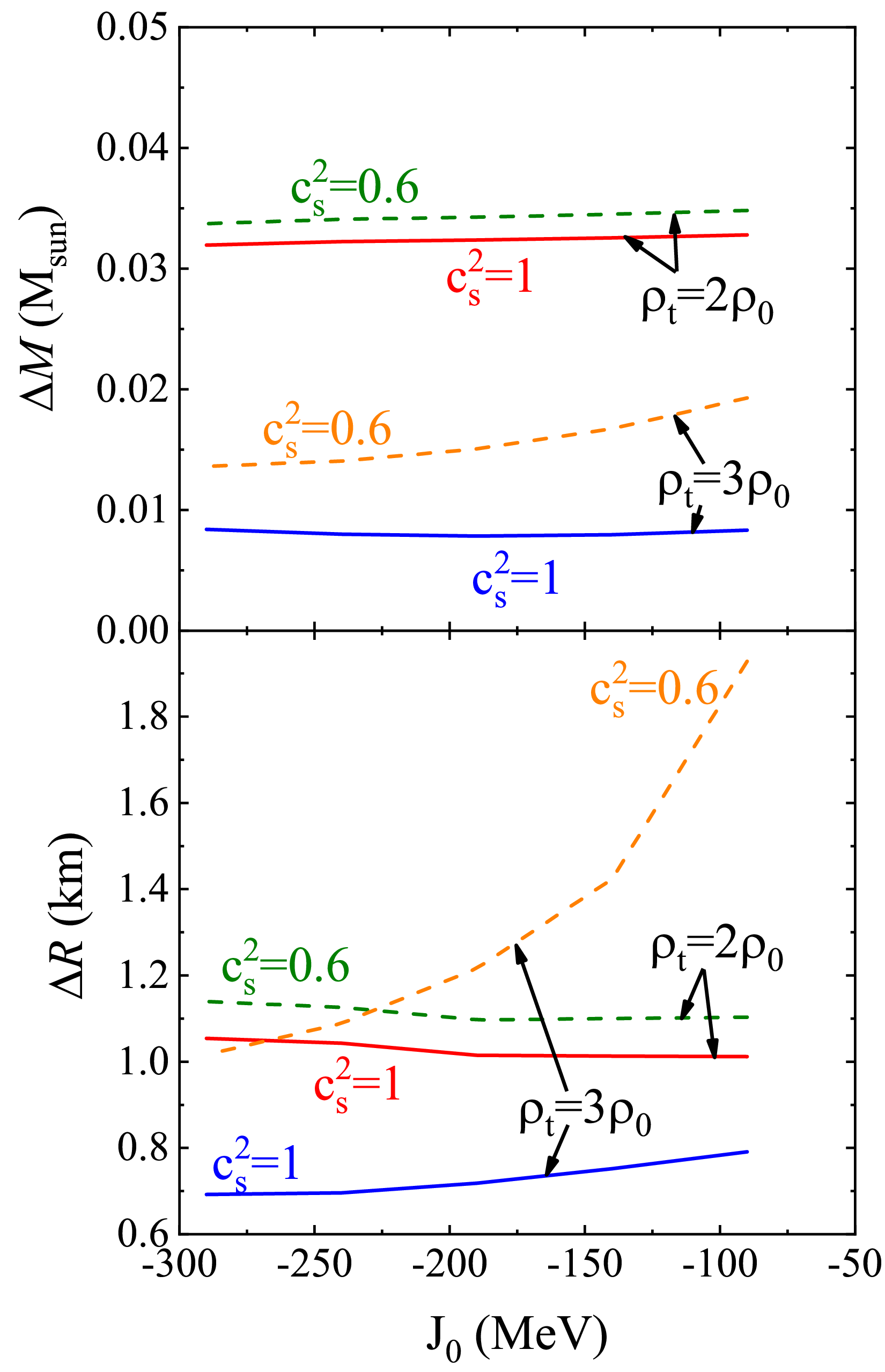}
  }
  \caption{The twin star mass range ${\rm \Delta} M$ (upper panel) and maximum radius separation ${\rm \Delta} R$ (lower panel) as functions of $J_0$ for different $\rho_t$ and $c_c^2$.}\label{FigJ0Delta}
\end{figure}

As indicated in Eq.~(\ref{E0-taylor}), the SNM EOS is determined by $E_0(\rho_0)$, $K_0$, and $J_0$. While the first two parameters are well-constrained by experiments, the uncertainty of $J_0$ remains relatively large, exerting a significant influence on the maximum mass of traditional NSs \cite{Xie24}. In previous studies, the $J_0$ is constrained to be approximately $-190\pm100$ MeV and we did not observe the formation of twin stars when nuclear matter parameters were fixed at their most probable values while varying the CSS parameters \cite{Zhang23}. However, Refs. \cite{LiJJ21,LiJJ23,LiJJ24} opted for a wide range of $J_0$ values, spanning from an upper limit of $1000$ MeV to a lower limit of $-600$ MeV. Consequently, it becomes imperative to meticulously examine the effects of $J_0$ on the formation of twin stars. Specifically, we need to ascertain whether twin stars can indeed form within $J_0=-190\pm100$ MeV.

The Fig. \ref{FigJ0} illustrates the effects of $J_0$ on the mass-radius relations when $\rho_t=2\rho_0, 3\rho_0$, and $c_s^2=0.6, 1$, respectively. Despite $J_0$ influencing the maximum mass of traditional NSs, we can see that the mass-radius curves are close to each other within the uncertainty of $J_0$. A smaller $J_0$ results in a softer EOS and a smaller mass at the transition point for both $\rho_t=2\rho_0$ and $\rho_t=3\rho_0$. The influence of $c_s^2$ on the maximum mass is reproduced again. For $\rho_t=2\rho_0$, the maximum masses converge for different $J_0$ values in hybrid stars. The uncertainty of $J_0$ introduces a maximum uncertainty of approximately 9\% to the mass at a fixed radius for both $c_s^2=1$ and $c_s^2=0.6$. As $\rho_t$ increases to $3\rho_0$, the effects of $J_0$ are slightly enhanced: the maximum uncertainty for the mass at a fixed radius does not converge and the deviation increases to about 15\%, and the maximum mass exhibits uncertainties of 3\% and 6\% for $c_s^2=1$ and $c_s^2=0.6$, respectively. The contribution of $J_0$ to the mass of traditional NSs is significantly mitigated by the relatively low transition density, given that the contribution from $J_0$ to the SNM EOS typically begins at around $2-3\rho_0$. However, larger uncertainties in $J_0$ might manifest notable effects on the mass-radius relations, thereby the formation of twin stars.

From Fig. \ref{FigJ0}, it's also apparent that all curves exhibit the twin star phenomenon. To further elucidate the effects of $J_0$ on the formation of twin stars, the twin star mass range ${\rm \Delta} M$ (upper panel) and maximum radius separation ${\rm \Delta} R$ (lower panel) as functions of $J_0$ for different $\rho_t$ and $c_s^2$ are depicted in Fig. \ref{FigJ0Delta}. The nearly horizontal lines indicate that $J_0$ exerts little influence on the formation of twin stars. Moreover, it's evident that increasing $c_s^2$ from 0.6 to 1 only marginally decreases the mass range of twin stars by less than 0.005 M$_\odot$, rendering the formation of twin stars slightly more challenging. Conversely, $\rho_t$ exhibits pronounced effects by significantly reducing the mass range of twin stars and making their formation substantially more difficult.

As shown in the lower panel of Fig. \ref{FigJ0Delta}, ${\rm \Delta} R$ remains relatively constant below about 1.2 km for different values of $\rho_t$ and $c_s^2$ , with the exception of $\rho_t=3\rho_0$ and $c_s^2=0.6$ that lead to a higher value of ${\rm \Delta} R$ approaching 2.0 km. Thus, compared to the current precision of radius measurements, twin stars become easier to detect if the phase transition occurs at a higher transition density and the EOS of quark matter is softer. The maximum ${\rm \Delta} R$ observed is approximately 1.95 km. Distinguishing them is beyond the current capabilities of present X-ray observatories and gravitational wave detectors. Similar results can be found in the following discussions for other parameters. However, future X-ray pulse profile observatories \cite{eXTP,STROBE-X} or gravitational wave detectors \cite{Hild,Sathyaprakash:2012jk,Evans:2021gyd}
are expected to measure the radii of neutron stars with precision to less than 0.1 km \cite{Chatziioannou:2021tdi,Pacilio:2021jmq,Finstad:2022oni,Bandopadhyay:2024zrr,Walker:2024loo}. Such high precision radius measurements certainly will help identify twin stars.

It's worth noting that the uncertainty range of $J_0$ is not chosen arbitrarily in the present manuscript. Previous studies by us and others using both forward modelings and/or Bayesian inferences have constrained $J_0$ to lie within the range of $-290$ to $90$ MeV based on analyses of available neutron star observational data \cite{Zhang19,Xie19,Xie20} as well as nuclear collective flow in relativistic heavy-ion collisions \cite{Xie21JPG}. Values outside this range would be inconsistent with the available data from both terrestrial experiments and astrophysical observations. Similarly, $E_0(\rho_0)$ and $K_0$ are also well constrained, and it is unnecessary to consider values beyond the current constraints for these parameters. Thus,  the effects of $E_0(\rho_0)$ and $K_0$ on twin star formation are not depicted here, given their relatively tight constraints compared to $J_0$, and they have little impact on the formation of twin stars (As an example, the effects of well constrained $E_{\rm sym}(\rho_0)$ on the formation of twin star is shown in the following discussions). Therefore, we can conclude that the EOS of SNM does not significantly influence the formation of twin stars.

\subsection{The effects of symmetry energy on twin stars}

\begin{figure}[ht]
  \centering
   \resizebox{0.45\textwidth}{!}{
  \includegraphics{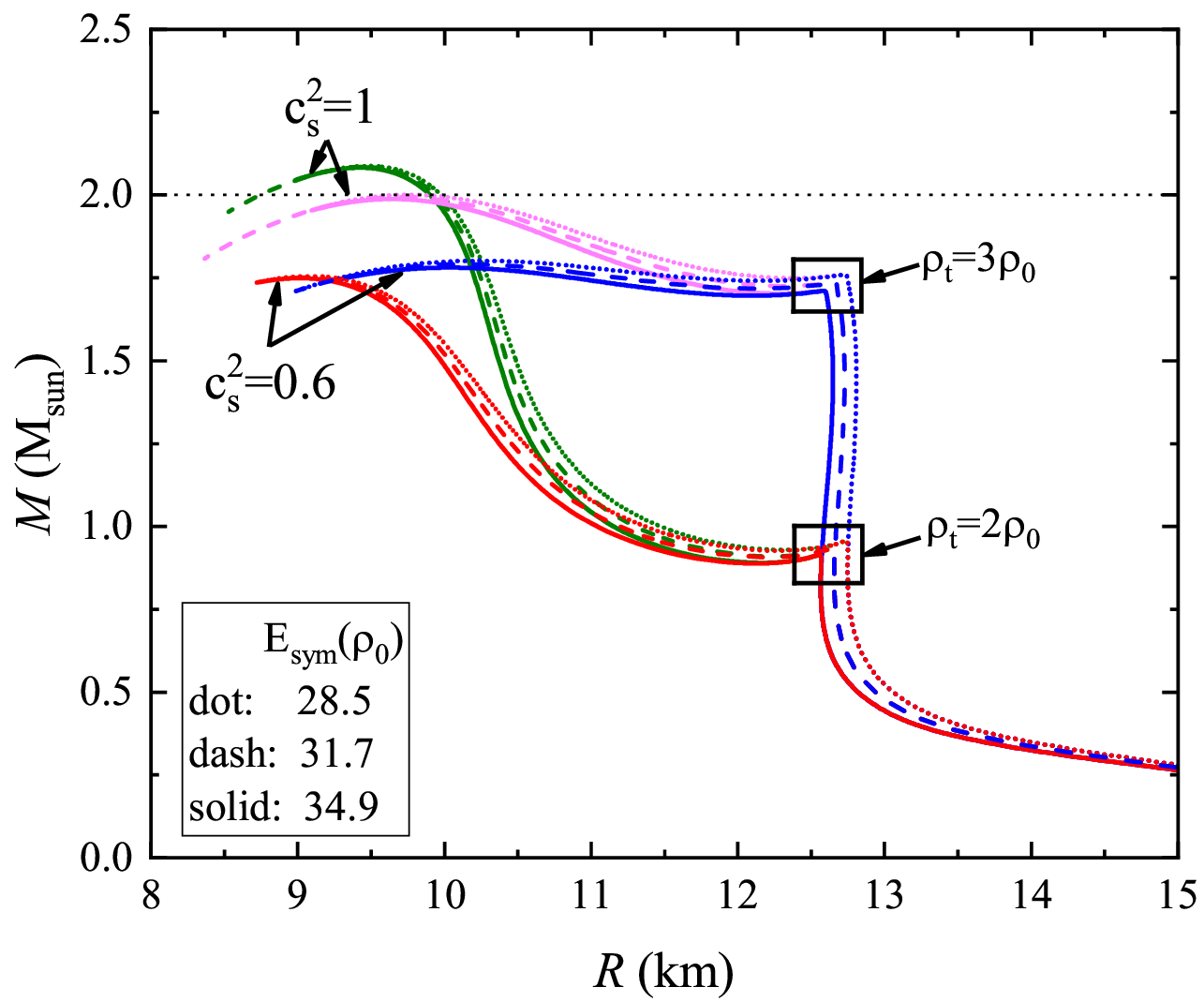}
  }
  \caption{Same as Fig. \ref{FigJ0} but for the symmetry energy at saturation density $E_{\rm sym}(\rho_0)$.}\label{FigEsym}
\end{figure}

\begin{figure}[ht]
  \centering
   \resizebox{0.4\textwidth}{!}{
  \includegraphics{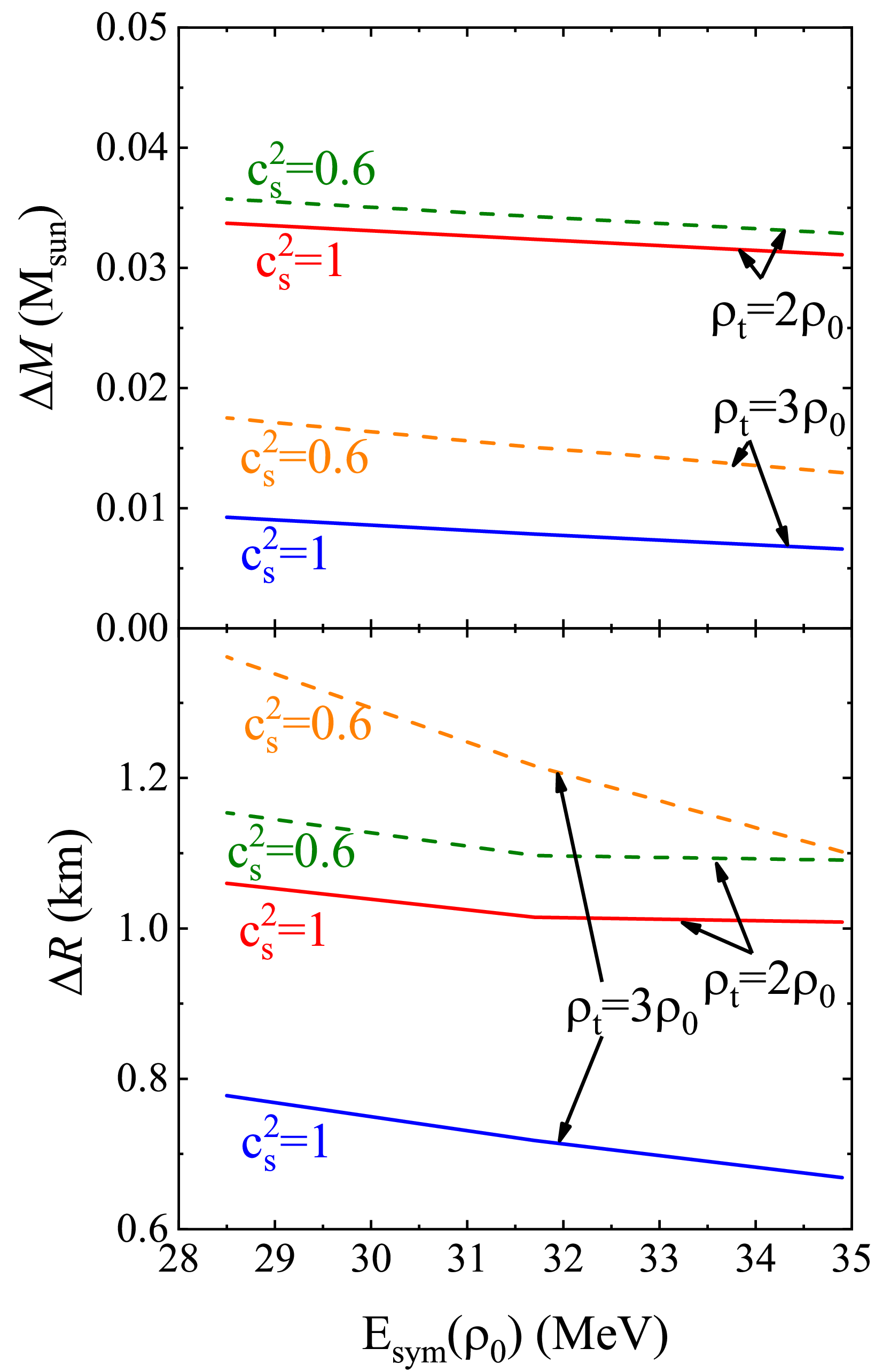}
  }
  \caption{Same as Fig. \ref{FigJ0Delta} but for the symmetry energy at saturation density $E_{\rm sym}(\rho_0)$.}\label{FigEsymDelta}
\end{figure}

It is well known that nuclear symmetry energy plays important roles in determining several properties of NSs, see e.g. Ref. \cite{Esym} for comprehensive reviews.
In particular, it is strongly correlated with the radii of canonical NSs with masses around 1.4 M$_\odot$, see, e.g., Refs.\cite{Andrew,Li06,James14}. Peculiarly, its curvature $K_{\rm sym}$ is found to be even more important than $L$ in determining the radius of canonical NSs \cite{Richter23}. As depicted in Fig. \ref{Figcss} and \ref{FigJ0}, twin stars can appear at any mass (as low as 0.5 M$_\odot$ and as high as 2 M$_\odot$) by simply varying the transition density, suggesting that the symmetry energy is likely to influence the formation of twin stars. The influence of $L$ on the formation of twin stars is presented in Ref. \cite{LiJJ21,LiJJ24}, indicating that greater values of $L$ facilitate the formation of twin stars. However, the effects of other symmetry energy parameters are seldom discussed.

In Fig. \ref{FigEsym}, the effects of $E_{\rm sym}(\rho_0)$ on the mass-radius relations are depicted for $\rho_t=2\rho_0, 3\rho_0$, and $c_s^2=0.6, 1$, respectively. Given that $E_{\rm sym}(\rho_0)$ is tightly constrained to be $31.7\pm3.2$ MeV, all the mass-radius curves appear nearly identical when varying the CSS parameters. However, in comparison to $J_0$, the smaller uncertainty of $E_{\rm sym}(\rho_0)$ results in a clearer decrease in ${\rm \Delta} M$ with increasing $E_{\rm sym}(\rho_0)$ in the upper panel of Fig. \ref{FigEsymDelta}, although the decrease remains subtle. More pronounced effects would be anticipated if $E_{\rm sym}(\rho_0)$ were less constrained.

We can also observe from the lower panel of Fig. \ref{FigEsymDelta} that an increase in $E_{\rm sym}(\rho_0)$ slightly complicates the identification of the twin star phenomenon and $c_s^2$ plays the most important role in determining the ${\rm \Delta} R$, especially for larger $\rho_t$. The maximum ${\rm \Delta} R=1.28$ km is still beyond the current capabilities of present X-ray observatories and gravitational wave detectors.

\begin{figure}[ht]
  \centering
   \resizebox{0.45\textwidth}{!}{
  \includegraphics{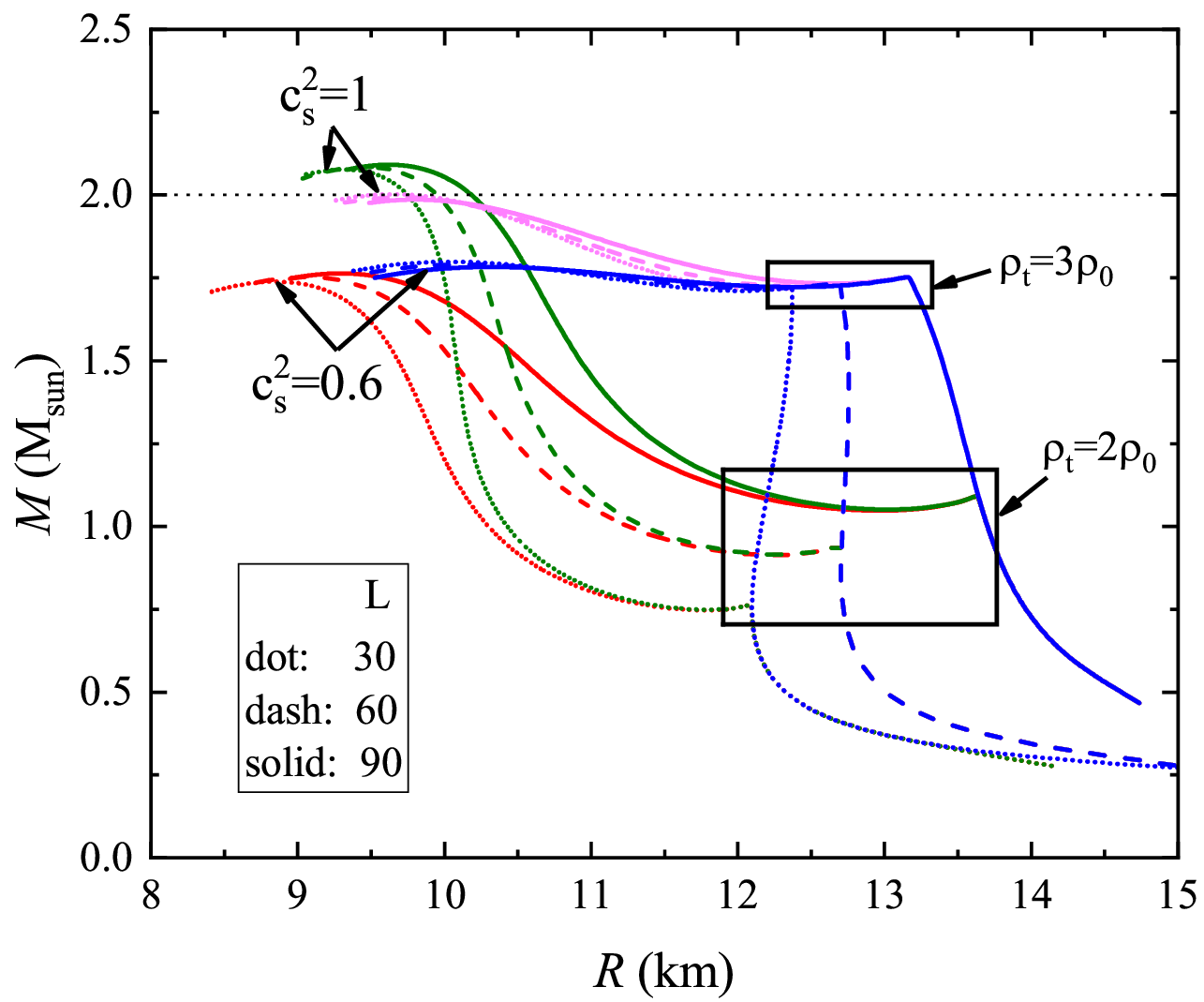}
  }
  \caption{Same as Fig. \ref{FigJ0} but for slope of symmetry energy $L$.}\label{FigL}
\end{figure}

As the slope of symmetry energy $L$ has larger uncertainty compared to $E_{\rm sym}(\rho_0)$ and it is sensitive to the radius of a canonical NS, $L$ should affect the formation of a twin star apparently. In Fig. \ref{FigL}, its effects are illustrated. For $\rho_t=2\rho_0$, the mass-radius curves exhibit distinct behavior, with mass (radius) increasing from 0.76 M$_\odot$ (12.10 km) to 1.09 M$_\odot$ (13.64 km) at the transition points as $L$ varies from 30 to 90 MeV, respectively. Although each mass-radius curve maintains separation, they gradually converge as they approach maximum masses of 1.75 and 2.09 M$_\odot$ for $c_s^2=0.6$ and $1$, respectively. However, the effects of $L$ are nearly imperceptible in Fig. \ref{FigL} for $\rho_t=3\rho_0$, as all curves merge together and only exhibit slight separation around 11-12 km, where the twin star phenomenon appears. This is because $L$ primarily influences the radius of canonical NSs but has little effect on the maximum mass (e.g., Fig. 3 in Ref. \cite{Zhang19a}). The contributions from $K_{\rm sym}$ and $J_{\rm sym}$ weaken the impact of $L$, and the $\delta^2$ term in the Eq. (\ref{Erho}) suppresses the contribution of the symmetry energy. Moreover, the relatively high transition density of $\rho_t=3\rho_0$ results in nearly identical transition masses around 1.75 M$_\odot$. The high transition density and energy density gap further attenuate the effects of $L$.

\begin{figure}[ht]
  \centering
   \resizebox{0.4\textwidth}{!}{
  \includegraphics{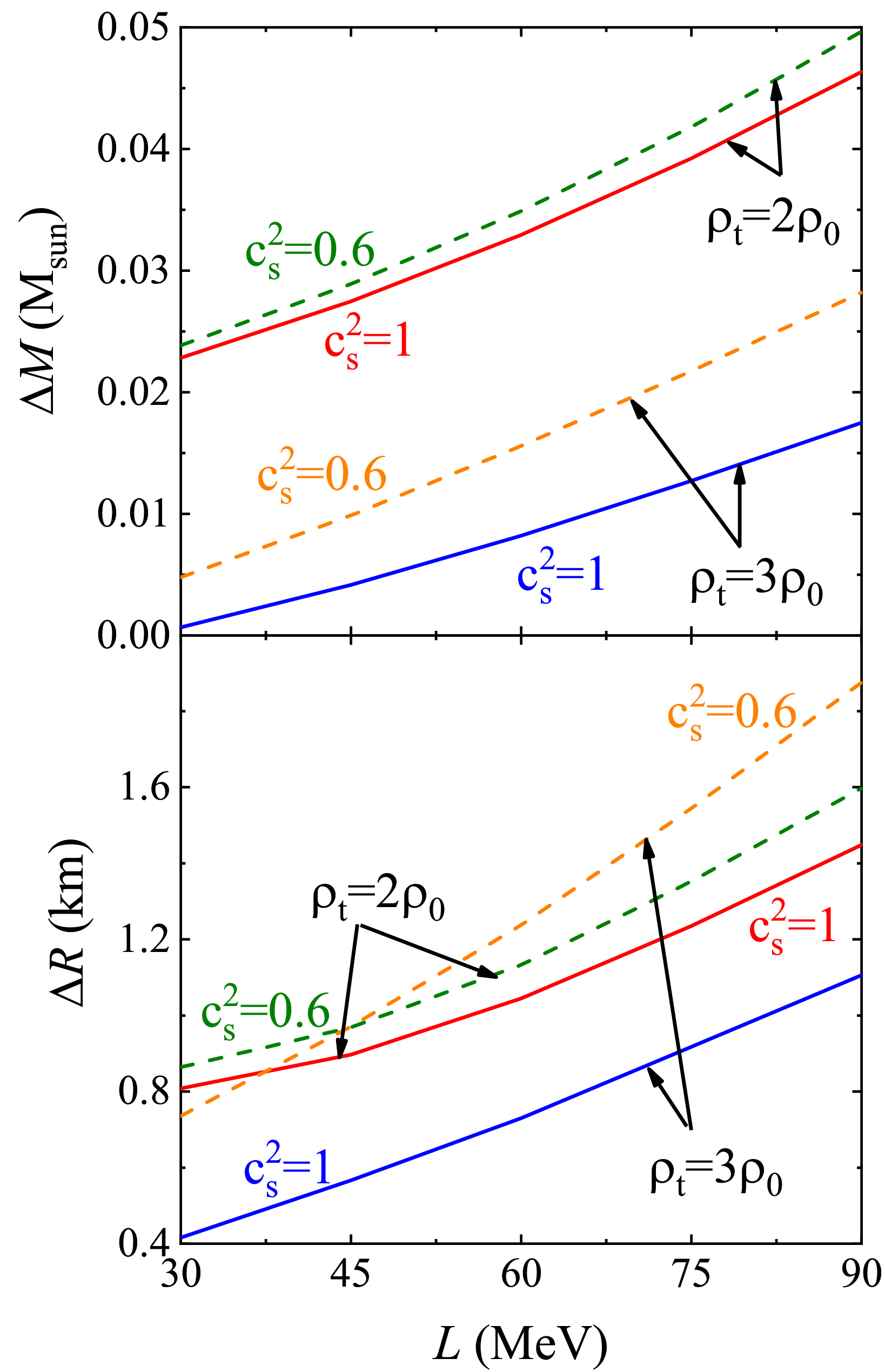}
  }
  \caption{Same as Fig. \ref{FigJ0Delta} but for slope of symmetry energy $L$.}\label{FigLDelta}
\end{figure}

To better illustrate the effects of $L$ on the formation of twin stars, the twin star mass range ${\rm \Delta} M$ and maximum radius separation ${\rm \Delta} R$ as functions of $L$ are depicted in Fig. \ref{FigLDelta}. It's evident that the mass range ${\rm \Delta} M$ increases with increasing $L$ regardless of the choice of CSS parameters. Since the mass range ${\rm \Delta} M$ of twin stars is typically smaller than 0.05 M$_\odot$ when ${\rm \Delta}\varepsilon=300$ MeV$\cdot$fm$^{-3}$, the effects of $L$ cannot be clearly discerned from the mass-radius relations alone. Instead, the ${\rm \Delta} M$ vs. $L$ relation should be examined. The twin star phenomenon would only disappear when $\rho_t=3\rho_0$, $c_s^2=1$, and $L=29$ MeV close to the lower limit of the uncertainty of $L$.

\begin{figure}[ht]
  \centering
   \resizebox{0.45\textwidth}{!}{
  \includegraphics{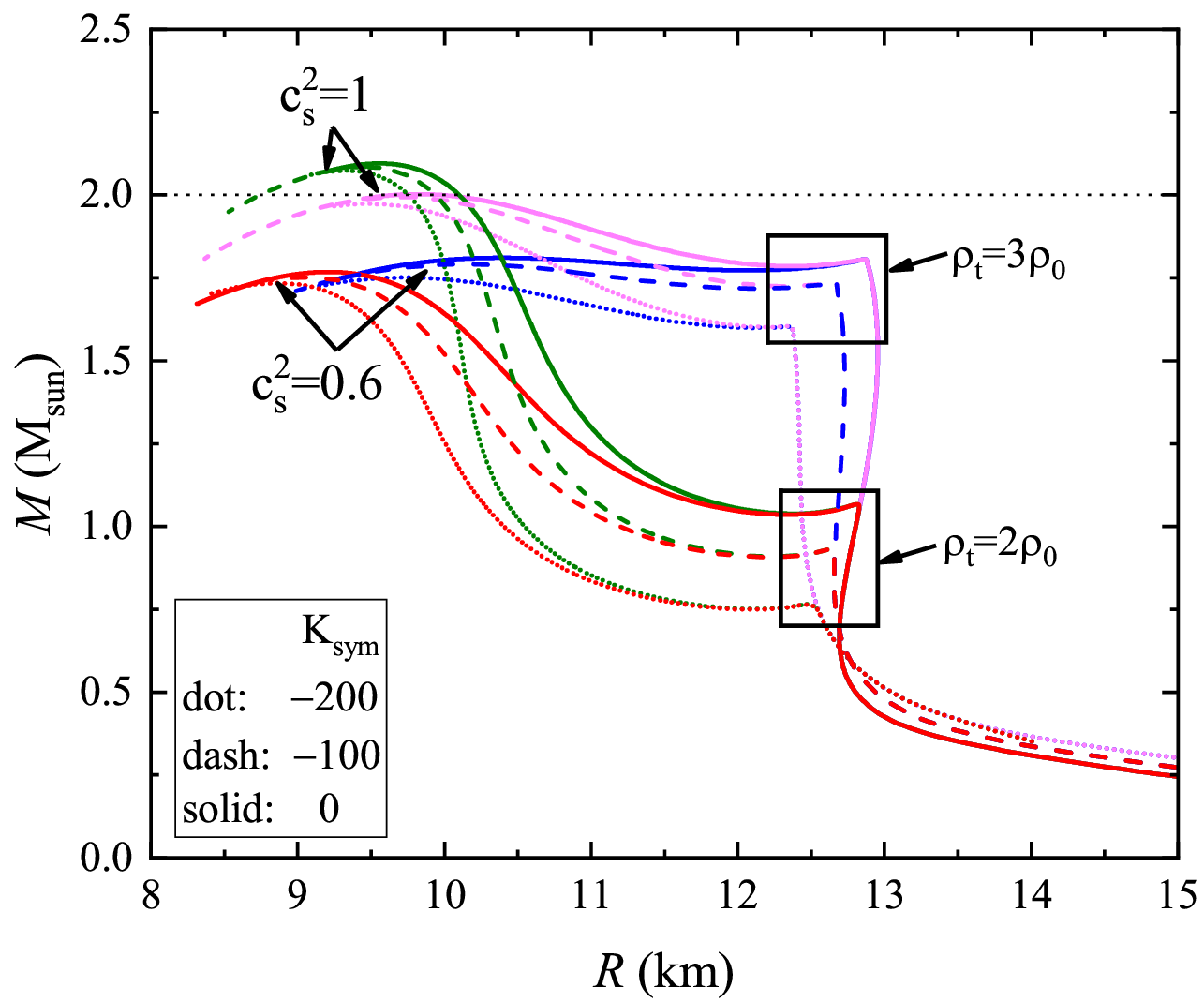}
  }
  \caption{Same as Fig. \ref{FigJ0} but for curvature of symmetry energy $K_{\rm sym}$.}\label{FigKsym}
\end{figure}

\begin{figure}[ht]
  \centering
   \resizebox{0.4\textwidth}{!}{
  \includegraphics{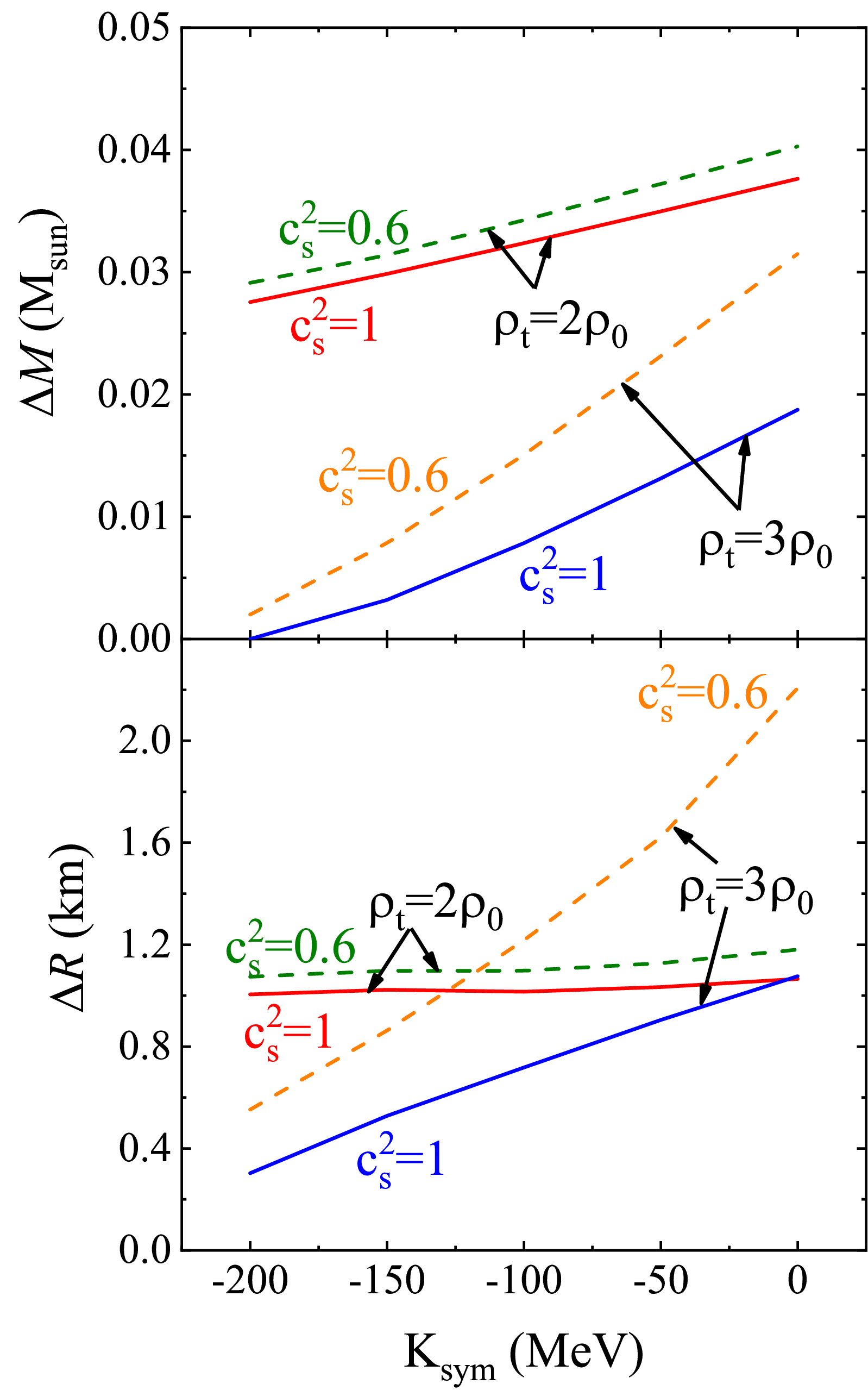}
  }
  \caption{Same as Fig. \ref{FigJ0Delta} but for curvature of symmetry energy $K_{\rm sym}$.}\label{FigKsymDelta}
\end{figure}

As a parameter affecting the EOS above about 2$\rho_0$, the effects of $K_{\rm sym}$ on the mass-radius relations are depicted in Fig. \ref{FigKsym}. Compared to the effects of $L$, the mass-radius relations before the phase transition are more crowded, but the transition mass differs for different $K_{\rm sym}$ values regardless of whether $\rho_t=2\rho_0$ or $\rho_t=3\rho_0$ is used. The trends of the mass-radius curves resemble the results shown in Fig. \ref{FigL} for $\rho_t=2\rho_0$, but noticeable effects emerge after the phase transition for $\rho_t=3\rho_0$. It is also demonstrated that the twin star phenomenon occurs for all the curves. To illustrate the effects of $K_{\rm sym}$ more clearly, the twin star mass range ${\rm \Delta} M$ and maximum radius separation ${\rm \Delta} R$ as functions of $K_{\rm sym}$ are presented in Fig. \ref{FigKsymDelta}. We can observe that $K_{\rm sym}$ exhibits almost the same impact as $L$ within their current uncertainties.

\begin{figure}[ht]
  \centering
   \resizebox{0.45\textwidth}{!}{
  \includegraphics{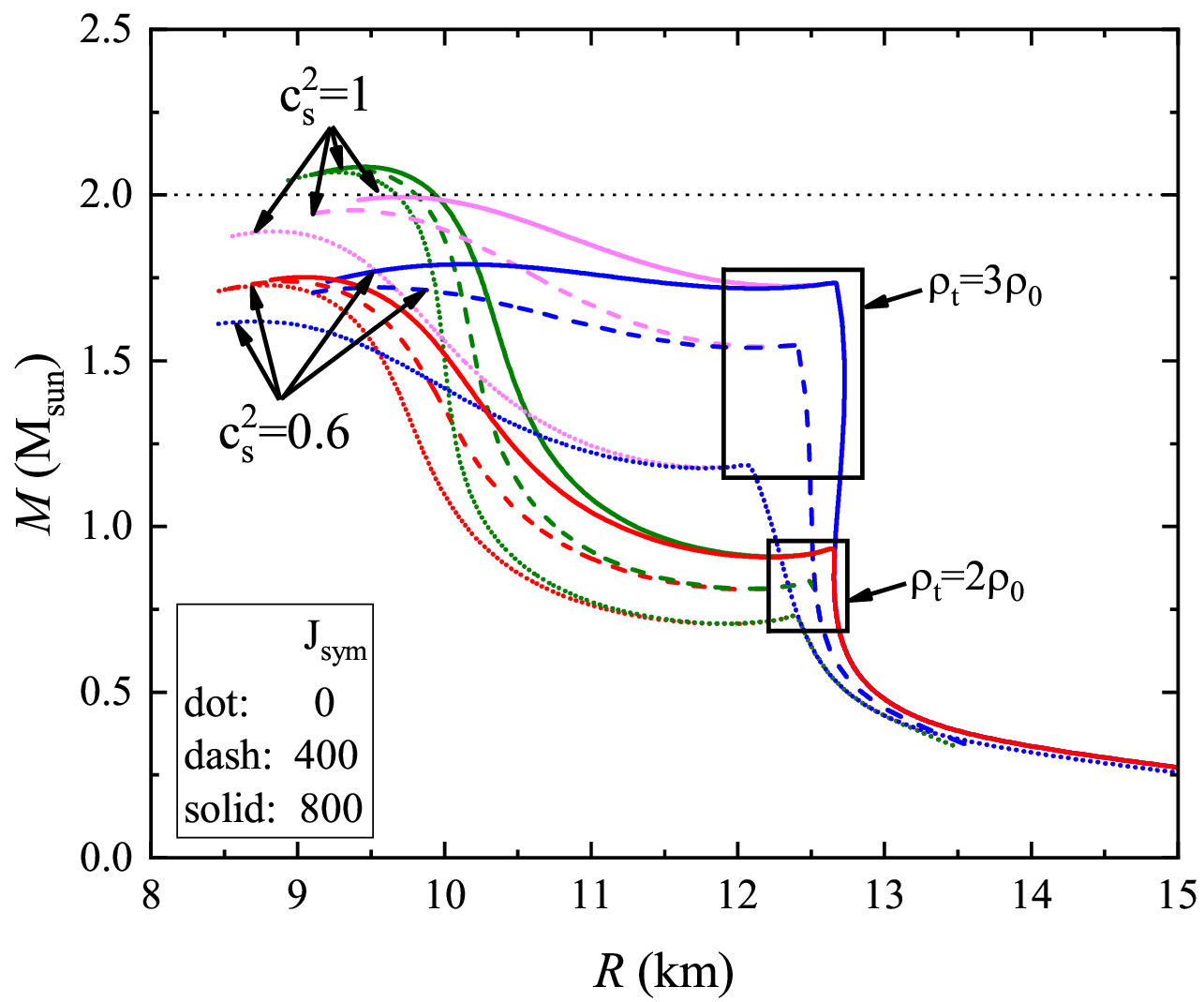}
  }
  \caption{Same as Fig. \ref{FigJ0} but for skewness of symmetry energy $J_{\rm sym}$.}\label{FigJsym}
\end{figure}

\begin{figure}[ht]
  \centering
   \resizebox{0.4\textwidth}{!}{
  \includegraphics{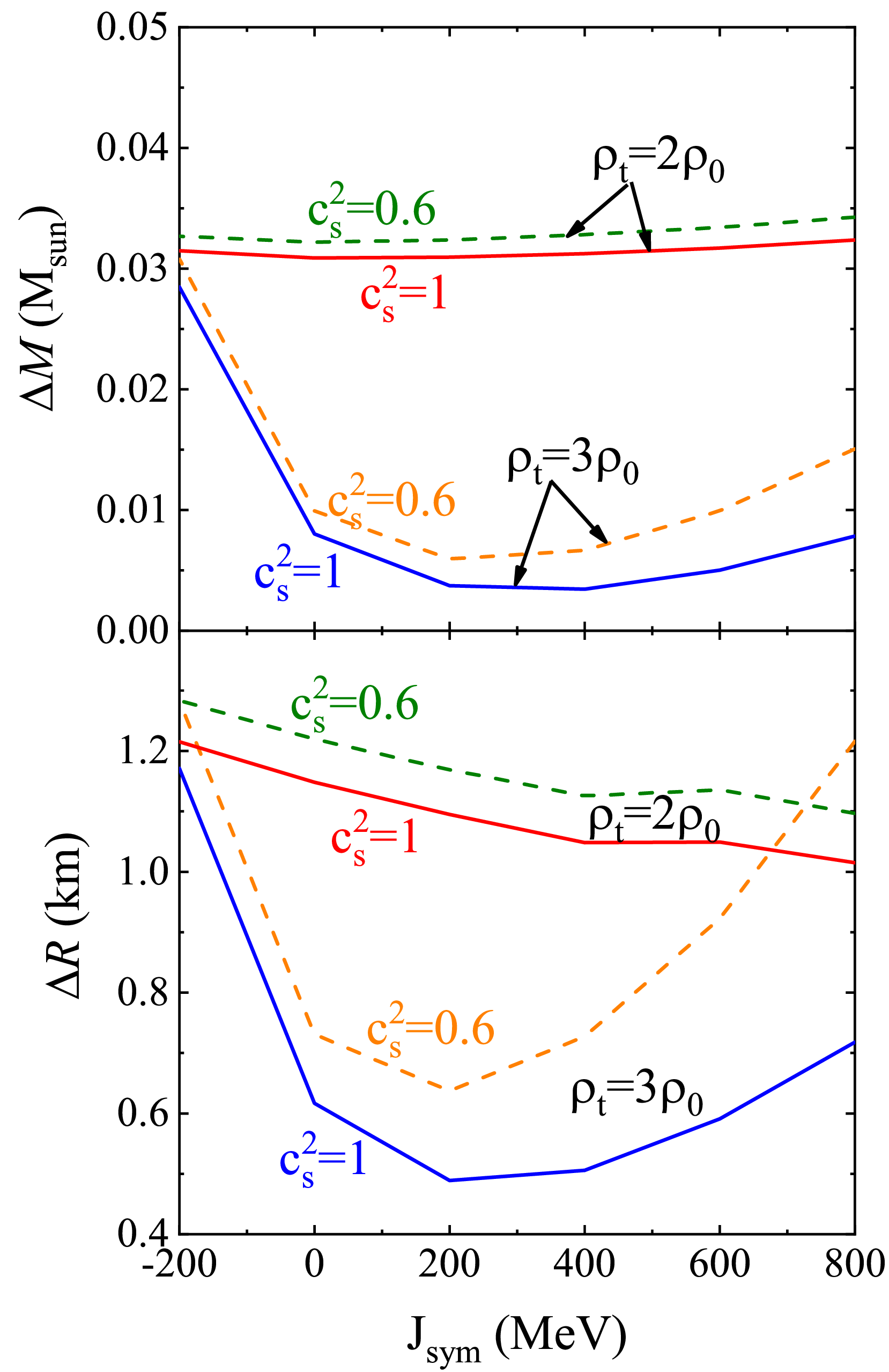}
  }
  \caption{Same as Fig. \ref{FigJ0Delta} but for skewness of symmetry energy $J_{\rm sym}$.}\label{FigJsymDelat}
\end{figure}

The skewness $J_{\rm sym}$ of symmetry energy is crucial in determining the EOS of neutron-rich nucleonic matter at high densities above about $3\rho_0$. Its impact on the mass-radius relations is depicted in Fig. \ref{FigJsym}. Comparing this figure with those of $L$ and $K_{\rm sym}$ in Figs. \ref{FigL} and \ref{FigKsym} reveals that though the effects of $L$, $K_{\rm sym}$, and $J_{\rm sym}$ on the transition mass or radius decrease successively for $\rho_t=2\rho_0$, they increase for $\rho_t=3\rho_0$. A higher transition density enhances the influence of parameters associated with high-density properties of nuclear matter, whereas a lower transition density suppresses it.

\begin{figure*}[htb!]
  \centering
   \resizebox{0.7\textwidth}{!}{
  \includegraphics{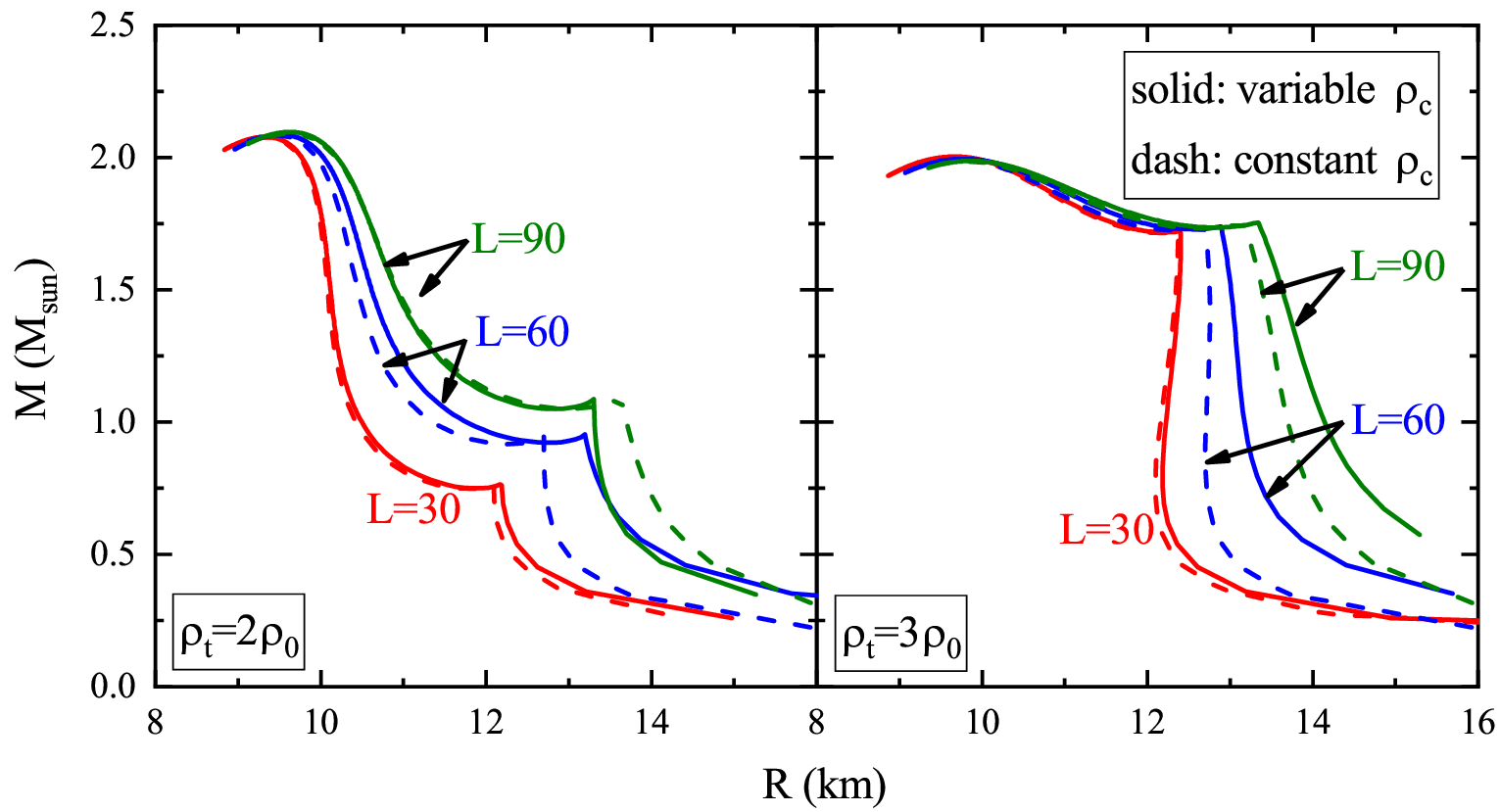}
  }
  \caption{The effects of a constant $\rho_c=0.08$ fm$^{-3}$ (solid lines) and variable $\rho_c$ calculated from Eq.~(\ref{Kmu}) (dashed lines) on the mass-radius relations for $\rho_t=2\rho_0$ (left panel) and $\rho_t=3\rho_0$ (right panel) with three representative values of $L=30$, $60$, and $90$ MeV, respectively. }\label{FigrhoL}
\end{figure*}

The twin star mass range ${\rm \Delta} M$ and maximum radius separation ${\rm \Delta} R$ as functions of $J_{\rm sym}$ are illustrated in Fig. \ref{FigJsymDelat}. For $\rho_t=2\rho_0$, the horizontal lines suggest that $J_{\rm sym}$ has little impact on the formation of twin stars. Similar trends are observed for $\rho_t=3\rho_0$ when $J_{\rm sym}\geq0$. This is because $J_{\rm sym}$ significantly affects the radii of massive NSs but has only a slight effect on their masses (see, e.g., Fig. 2 in Ref. \cite{Xie24}). However, for $J_{\rm sym}<0$, the high-density symmetry energy becomes extremely soft, even decreasing with increasing density. Consequently, the isospin asymmetry ${\rm \Delta}$ at $\beta$-equilibrium increases due to the well-known isospin fraction phenomenon \cite{LCK}, accentuating the contribution of the symmetry energy to the EOS according to Eq. (\ref{Erho}). More detailed discussions on the astrophysical impact of the super-soft symmetry energy at high densities can be found in Ref. \cite{EPJA-review}.
As a result, a softer high-density EOS is obtained, and the corresponding maximum mass is much smaller than 2 M$_\odot$, potentially leading to its exclusion even without considering phase transitions. Thus, it can be concluded that $J_{\rm sym}$ does not affect the formation of twin stars.

\subsection{The effects of crust-core transition density on twin stars}

In the above discussions, we have fixed the crust-core transition density at its fiducial value $\rho_c=0.08$ fm$^{-3}$ widely used in the literature.
This choice excludes its potential influence on the formation of twin stars and thus enables us to focus on studying how the EOS of SNM and the symmetry energy parameters may affect the formation of twin stars. However, since both $\rho_c$ and $E_{\rm sym}$ can influence the radius of a neutron star, and consequently affect the formation of twin stars, their combined effects on the formation of twin stars should be carefully considered.

As an example, we compare in Fig. \ref{FigrhoL} the results obtained by using a constant $\rho_c=0.08$ fm$^{-3}$ (solid lines) and variable $\rho_c$ calculated self-consistently from Eq.~(\ref{Kmu}) (dashed lines) on the mass-radius relations for $\rho_t=2\rho_0$ (left panel) and $\rho_t=3\rho_0$ (right panel) with three representative values of $L=30$, $60$, and $90$ MeV, respectively. We found that the calculated variable $\rho_c$ is generally larger than the fiducial value for the EOS parameters considered. It results in noticeably larger radii but nearly constant mass when the phase transition from hadronic matter to quark matter occurs. This is because $\rho_c$ increases with $L$ when $K_{\rm sym}$ is fixed as $-100$ MeV, as shown in Fig. 3 of Ref. \cite{Zhang18}. Specifically, the crust-core transition density increases from $0.0928$ fm$^{-3}$ for $L=30$ MeV to $0.1024$ fm$^{-3}$ and $0.1312$ fm$^{-3}$ for $L=60$ and $90$ MeV, respectively. Since $\rho_c$ for $L=30$ MeV is quite close to $0.08$ fm$^{-3}$, the red lines almost merge together. Similar results are observed when varying $K_{\rm sym}$ and $J_{\rm sym}$. However, due to the limitations of our present model framework, we are unable to analyze the effects of all parameters on the formation of twin star simultaneously. Nevertheless, we emphasize that the crust-core transition density should be treated with care when studying the physics related to the radii of neutron stars.

\section{Summary and outlook}

In summary, within a meta-model for hybrid NS EOSs, we investigated the viability of twin stars and how the EOS parameters and crust-core transition density influence their formation, instead of how the current NS observations may constrain the properties of twin stars. We found that the existence of twin stars remains inconclusive, as we first need to precisely determine numerous EOS parameters characterizing nuclear matter, quark matter, and the phase transition between them. The EOS of SNM exhibits minimal influence on the formation of twin stars. In contrast, the nuclear symmetry energy, particularly its slope $L$ and curvature $K_{\rm sym}$, significantly impacts the formation of twin stars. Moreover, the largest mass range for twin stars to coexist is found to be about ${\rm \Delta} M\approx0.05$ M$_\odot$, indicating that the formation of twin stars is infrequent based on our current knowledge about NS EOS. However, the mass range can be enhanced if the ${\rm \Delta}\varepsilon$ values are larger than 300 MeV.

In addition, we found that the largest radius range ${\rm \Delta} R$ separating twin stars is smaller than 2.0 km. Twin stars have not been observed partially because their radius separation ${\rm \Delta} R$ predicted by most acceptable EOSs is less than $2\sigma_{\rm obs}\approx$2 to 4 km of currently available neutron star observatories. However, future X-ray pulse profile observatories or gravitational wave detectors may be able to achieve the necessary precision to distinguish twin stars. By examining the effects of crust-core transition density on the formation of twin star, we found indications that the crust-core transition density should be treated with care when studying the physics related to the radii of neutron stars.

The CSS model assumes that the phase transition from nuclear matter to quark matter is first-order, and the speed of sound of quark matter is constant. We found that the speed of sound has a minimal impact compared to the other two CSS parameters. However, since the transition densit of pressure and chemical potential, the effects of the speed of sound on the formation of twin stars would become more significant if the correlations among the CSS parameters are taken into account. In addition, we also note that Ref. \cite{Essick24} proposedy $\rho_t$ and the gap in energy density $\Delta\varepsilon$ are determined by both the hadronic and quark matter EOS in the plane very recently that twin stars could be formed without considering the first-order phase transition when the $c_s^2$ monotonically increases with energy density. The second stable branch on the mass-radius curve might emerge at super-high densities.

Finally, all the discussions above are for neutron star matter at zero temperature. Interestingly, based on the phase diagram (e.g., Fig.~1 from Ref. \cite{Baym18}), twin stars could also be formed at finite temperatures, leading to the concept of thermal twin stars \cite{Hempel16}. More specifically, they can be formed at temperatures of $T = 30$ MeV and above, without altering properties of the phase transition or quark matter \cite{Carlomagno24}. Further studies are needed to find out how the finite temperature may alter what we have found above in this work.

\section*{Acknowledgement}
We would like to thank Bao-Jun Cai, Xavier Grundler and Wen-Jie Xie for helpful discussions. BAL is supported in part by the U.S. Department of Energy, Office of Science, under Award Number DE-SC0013702, the CUSTIPEN (China-U.S. Theory Institute for Physics with Exotic Nuclei) under the US Department of Energy Grant No. DE-SC0009971. NBZ is supported in part by the National Natural Science Foundation of China under Grant No. 12375120, the Zhishan Young Scholar of Southeast University under Grant No. 2242024RCB0013 and the Start-up Research Fund of Southeast University under Grant No. RF1028623060.
%

\begin{thebibliography}{1}
\bibitem{Alford13} M.G. Alford, S. Han, M. Prakash, Phys. Rev. D \textbf{88}, 083013 (2013)
\bibitem{Bardeen66} J.M. Bardeen, K.S. Thorne, D.W. Meltzer, Astrophys. J. \textbf{145}, 505 (1966)
\bibitem{Harrison65} B. K. Harrison, K. S. Thorne, M. Wakano, J. A. Wheeler, {\it Gravitation theory and gravitational collapse}, Chicago: University of Chicago Press (1965)
\bibitem{Gerlach68} U.H. Gerlach, Phys. Rev. \textbf{172}, 1325 (1968)
\bibitem{Kampfer81} K\"{a}mpfer, J. Phys. A: Math. Gen. \textbf{14}, L471, (1981)
\bibitem{Kampfer81b} K\"{a}mpfer, Phys. Lett. B \textbf{101}, 366 (1981)
\bibitem{Glendenning20} N.K. Glendenning, C. Kettner, Astron. Astrophys. \textbf{353}, L9 (2000)
\bibitem{Schertler20} K. Schertler, C. Greiner, J. Schaffner-Bielich, M.H. Thoma, Nucl. Phys. A \textbf{677}, 463 (2000)
\bibitem{Miller19} M.C. Miller {\it et al.}, Astrophys. J. Lett. \textbf{887}, L24 (2019)
\bibitem{Riley19} T.E. Riley {\it et al.}, Astrophys. J. Lett. \textbf{887}, L21 (2021)
\bibitem{Miller21} M.C. Miller {\it et al.}, Astrophys. J. Lett. \textbf{918}, L28 (2021)
\bibitem{Riley21} T.E. Riley {\it et al.}, Astrophys. J. Lett. \textbf{918}, L27 (2021)
\bibitem{LIGO18} B.P. Abbott {\it et al.}, Phys. Rev. Lett. \textbf{121}, 161101 (2018)
\bibitem{Alford16} M. G. Alford and S. Han, Euro. Phys. J. A \textbf{52}, 66 (2016)
\bibitem{Ranea16} I.F. Ranea-Sandoval, S. Han, M.G. Orsaria, G.A. Contrera, F. Weber, M.G. Alford, Phys. Rev. C \textbf{93}, 045812 (2016)
\bibitem{Montana19} G. Montana, L. Tolos, M. Hanauske, L. Rezzolla, Phys. Rev. D \textbf{99}, 103009 (2019)
\bibitem{Espino22} P. Espino, V. Paschalidis, Phys. Rev. D \textbf{105}, 043014 (2022)
\bibitem{Christian20} Jan-Erik Christian, J. Schaffner-Bielich, Astrophys. J. \textbf{894}, L8 (2020)
\bibitem{Christian21} Jan-Erik Christian, J. Schaffner-Bielich, Phys. Rev. D \textbf{103}, 063042 (2021)
\bibitem{Christian22} Jan-Erik Christian, J. Schaffner-Bielich, Astrophys. J. \textbf{935}, 122 (2022)
\bibitem{Tsaloukidis23} Lazaros Tsaloukidis, P.S. Koliogiannis, A. Kanakis-Pegios, Ch. C. Moustakidis, Phys. Rev. D \textbf{107}, 023012 (2023)
\bibitem{LiJJ21} J.J. Li, A. Sedrakian, M. Alford, Phys. Rev. D \textbf{104}, L121302 (2021)
\bibitem{LiJJ23} J.J. Li, A. Sedrakian, M. Alford, Astrophysical J. \textbf{944}, 206 (2023)
\bibitem{LiJJ24} J.J. Li, A. Sedrakian, M. Alford, Astrophys. J. \textbf{967}, 116 (2024).
\bibitem{Gorda23} T. Gorda, K. Hebeler, A. Kurkela, A. Schwenk, A. Vuorinen, Astrophys. J. \textbf{955}, 100 (2023)
\bibitem{Sun23} H. Y. Sun and D. H. Wen, Phys. Rev. C \textbf{108}, 025801 (2023)
\bibitem{Pradhan23} B.K. Pradhan, D. Chatterjee, D.E. Alvarez-Castillo, 
Mon. Not. Roy. Astron. Soc. \textbf{531}, no.4, 4640-4655 (2024)
\bibitem{Imajo24} S. Imajo, A. Miyake, R. Kurihara, M. Tokunaga, K. Kindo, S. Horiuchi, F. Kagawa, 
Phys. Rev. Lett. {\bf 132}, 096601 (2024)
\bibitem{Carlomagno24}  J.P. Carlomagno, G.A. Contrera, A.G. Grunfeld, D. Blaschke, Phys. Rev. D \textbf{109}, 043050 (2024)
\bibitem{Xie21} W.J. Xie, B.A. Li, Phys. Rev. C \textbf{103}, 035802 (2021)
\bibitem{Zhang23} N.B. Zhang, B.A. Li, Phys. Rev. C \textbf{108}, 025803 (2023)
\bibitem{Zhang19} N.B. Zhang, B.A. Li, Euro. Phys. J. A \textbf{55}, 39 (2019)
\bibitem{Xie19} W.J. Xie, B.A. Li, Astrophys. J. \textbf{883}, 174 (2019)
\bibitem{Xie20} W.J. Xie, B.A. Li, Astrophys. J. \textbf{899}, 4 (2020)
\bibitem{Xie21JPG} W.J. Xie, B.A. Li, J. Phys. G \textbf{48}, 025110 (2021)
\bibitem{Wang} G. Gary Wang, S. Shan, J. Mech. Des., {\bf 129(4)}, 370 (2007)
\bibitem{Zhang18} N.B. Zhang, B.A. Li, and J. Xu, Astrophys. J. \textbf{859}, 90 (2018)
\bibitem{Zhang19a} N.B. Zhang, B.A. Li, J. Phys. G: Nucl. Part. Phys. \textbf{46}, 014002 (2019)
\bibitem{Zhang19b} N.B. Zhang, B.A. Li, Astrophys. J. \textbf{879}, 99 (2019)
\bibitem{Zhang2020} N.B. Zhang, B.A. Li, Astrophys. J. \textbf{883}, 61 (2020)
\bibitem{Zhang2021} N.B. Zhang, B.A. Li, Astrophys. J. \textbf{921}, 111 (2021)
\bibitem{Zhang22} N.B. Zhang and B.A. Li, Euro. Phys. J. A \textbf{59}, 86 (2023)
\bibitem{Xie24} W.J. Xie, B.A. Li, N.B. Zhang, Phys. Rev. D \textbf{110}, 043025 (2024)
\bibitem{Garg18} U. Garg, G. Col\`{o}, Prog. Part. Nucl. Phys. \textbf{101}, 55 (2018)
\bibitem{Shlomo06} S. Shlomo, V. M. Kolomietz, G. Col\`{o}, Euro. Phys. J. A \textbf{30}, 23 (2006)
\bibitem{Li13} B.A. Li, X. Han, Phys. Lett. B \textbf{727}, 276 (2013)
\bibitem{Oertel17} M. Oertel, M. Hempel, T. Kl\"{a}hn, S. Typel, Rev. Mod. Phys. \textbf{89}, 015007 (2017)
\bibitem{Li21} B.A. Li, B.J. Cai, W.J. Xie, N.B. Zhang, Universe \textbf{7}, 182 (2021)
\bibitem{Grams22} G. Grams, R. Somasundaram, J. Margueron, E. Khan, Phys. Rev. C \textbf{106}, 044305 (2022)
\bibitem{Margueron18} J. Margueron, R.H. Casali, F. Gulminelli, Phys. Rev. C \textbf{97}, 025806 (2018)
\bibitem{Mondal17} C. Mondal, B.K. Agrawal, J.N. De, S.K. Samaddar, M. Centelles, X. Vi\~{n}as, Phys. Rev. C \textbf{96}, 021302 (2017)
\bibitem{Somasundaram21} R. Somasundaram, C. Drischler, I. Tews, J. Margueron, Phys. Rev. C \textbf{103}, 045803 (2021)
\bibitem{Cai17} B.J. Cai, L.W. Chen, Nucl. Sci. Tech. \textbf{28}, 185 (2017)
\bibitem{Zhang17} N.B. Zhang, B.J. Cai, B.A. Li, W.G. Newton, J. Xu, , Nucl. Sci. Tech. \textbf{28}, 181 (2017)
\bibitem{Tews17} I. Tews, J.M. Lattimer, A. Ohnishi, E.E. Kolomeitsev, Astrophys. J. \textbf{848}, 105 (2017)
\bibitem{Oppenheimer39} J. Oppenheimer, G. Volkoff, Phys. Rev. \textbf{55}, 374 (1939)
\bibitem{Negele73} J.W. Negele, D. Vautherin, Nucl. Phys. A \textbf{207}, 298 (1973)
\bibitem{Baym71b} G. Baym, C.J. Pethick, P. Sutherland, Astrophys. J. \textbf{170}, 299 (1971)
\bibitem{Kubis04} S. Kubis, Phys. Rev. C \textbf{70}, 065804 (2004)
\bibitem{Kubis07} S. Kubis, Phys. Rev. C \textbf{76}, 025801 (2007)
\bibitem{Lattimer07} J. M Lattimer and M. Prakash, Phys. Rep. \textbf{442}, 109 (2007)
\bibitem{Chamel13} N. Chamel, A. Fantina, J. Pearson, S. Goriely, Astron. Astrophys. \textbf{553}, A22 (2013)
\bibitem{Zdunik13} J. Zdunik, P. Haensel, Astron. Astrophys. \textbf{551}, A61 (2013)
\bibitem{Agrawal10} B. Agrawal, Phys. Rev. D \textbf{81}, 023009 (2010)
\bibitem{Bonanno12} L. Bonanno, A. Sedrakian, Astron. Astrophys. \textbf{539}, A16 (2012)
\bibitem{Lastowiecki12} R. Lastowiecki, D. Blaschke, H. Grigorian, S. Typel, Acta Phys. Pol. B Proc. Suppl. \textbf{5}, 535 (2012)
\bibitem{Kurkela10} A. Kurkela, P. Romatschke, A. Vuorinen, B. Wu, arXiv:1006.4062 (2010)
\bibitem{Kurkela10b} A. Kurkela, P. Romatschke, A. Vuorinen, Phys. Rev. D \textbf{81}, 105021 (2010)
\bibitem{Traversi21} S. Traversi, P. Char, G. Pagliara, A. Drago, Astron. Astrophys. \textbf{660}, A62 (2022)
\bibitem{Traversi20} S. Traversi, P. Char, Astrophys. J. \textbf{905}, 9 (2020)
\bibitem{Alford15} M.G. Alford, G.F. Burgio, S. Han, G. Taranto, D. Zappal\`{a}, Phys. Rev. D \textbf{92}, 083002 (2015)
\bibitem{Ayriyan15} A. Ayriyan, D.E. Alvarez-Castillo, D. Blaschke, H. Grigorian, M. Sokolowski, Phys. Part. Nucl. \textbf{46}, 854 (2015)
\bibitem{Drischler22} C. Drischler, S. Han, J.M. Lattimer, M. Prakash, S. Reddy, T.Q. Zhao, Phys. Rev. C \textbf{103}, 045808 (2021)
\bibitem{Chatziioannou20} K. Chatziioannou, S. Han, Phys. Rev. D \textbf{101}, 044019 (2020)
\bibitem{Han20} S. Han, M. Prakash, Astrophys. J. \textbf{899}, 164 (2020)
\bibitem{Li22} A. Li, G.C. Yong, Y.X. Zhang, Phys. Rev. D \textbf{107}, 043005 (2023)
\bibitem{Miao20} Z.Q. Miao, A. Li, Z.Y. Zhu, S. Han, Astrophys. J. \textbf{904}, 103 (2020)
\bibitem{Backes21} B.C. Backes, E. Hafemann, I. Marzola, D.P. Menezes, J. Phys. G \textbf{48}, 055104 (2021)
\bibitem{Wen05} X.J. Wen, X.H. Zhong, G.X. Peng, P.N. Shen, P.Z. Ning, Phys. Rev. C \textbf{72}, 015204 (2005)
\bibitem{Albino24} M. Albino, T. Malik, M. Ferreira, C. Provid\^{e}ncia, Phys. Rev. D \textbf{110}, 083037 (2024)
\bibitem{Roy24}D.G. Roy, A. Venneti, T. Malik, S. Bhattacharya, S. Banik, Phys. Lett. B \textbf{859}, 139128 (2024)
\bibitem{Liang21} A. Li, Z.Q. Miao, S. Han, B. Zhang, Astrophys. J. \textbf{913}, 27 (2021)
\bibitem{Somasundaram22} R. Somasundaram, J. Margueron, Europhysics Letters \textbf{138}, 14002 (2022)
\bibitem{Seidov71} Z.F. Seidov, Sov. Astron. \textbf{15}, 347 (1971)
\bibitem{Schaeffer83} R. Schaeffer, L. Zdunik, P. Haensel, Astron. Astrophys. \textbf{126}, 121 (1983)
\bibitem{Lindblom98} L. Lindblom, Phys. Rev. D \textbf{58}, 024008 (1998)
\bibitem{Doroshenko22} V. Doroshenko, V. Suleimanov, G. P\"{u}hlhofer, A. Santangelo, Nature Astronomy \textbf{6}, 1444 (2022)
{\bibitem{Typel16} S. Typel, Eur. Phys. J. A \textbf{52}, 16 (2016)
\bibitem{Benic15} S. Beni\'{c}, D. Blaschke, D.E. Alvarez-Castillo, T. Fischer, S. Typel, Astron. Astrophys. \textbf{577}, A40 (2015)
\bibitem{Esym}Topical Issue on Nuclear Symmetry Energy, Eds. B.A. Li, A. Ramos, G. Verde, I. Vidana, The European Physical Journal A \textbf{50}, 9 (2014)
\bibitem{eXTP} S.N. Zhang, et~al., Sci. China Phys. Mech. Astron.  {\bf62}, 29502 (2019)
\bibitem{STROBE-X} P.S. Ray, et~al., STROBE-X: X-ray Timing and Spectroscopy on Dynamical Timescales from Microseconds to Years,arXiv:1903.03035
\bibitem{Hild} S.~Hild, S.~Chelkowski, A.~Freise, et~al., Class. Quant. Grav.  {\bf27}, 015003 (2010)
\bibitem{Sathyaprakash:2012jk} B.~Sathyaprakash, et~al., Class. Quant. Grav.  {\bf29}, 124013 (2012) [Erratum: Class. Quant. Grav. 30, 079501 (2013)].
\bibitem{Evans:2021gyd} M.~Evans, et~al., arXiv:2109.09882
\bibitem{Chatziioannou:2021tdi} K.~Chatziioannou, Phys. Rev. D {\bf105}, 084021 (2022)
\bibitem{Pacilio:2021jmq} C.~Pacilio, A.~Maselli, M.~Fasano, et~al., Phys. Rev. Lett.  {\bf128}, 101101 (2022)
\bibitem{Finstad:2022oni} D.~Finstad, L.V. White, D.A. Brown, Astrophys. J.  {\bf955}, 45 (2023)
\bibitem{Bandopadhyay:2024zrr} A.~Bandopadhyay, K.~Kacanja, R.~Somasundaram, et~al., Class. Quant. Grav. \textbf{41}, no.22, 225003 (2024)
\bibitem{Walker:2024loo} K.~Walker, R.~Smith, E.~Thrane, et~al., Phys. Rev. D \textbf{110}, no.4, 043013 (2024)
\bibitem{Andrew}A.W. Steiner, M. Prakash, J.M. Lattimer, P.J. Ellis, Phys. Rept. \textbf{411}, 325 (2005)
\bibitem{Li06} B. A. Li, A. W. Steiner, Phys. Lett. B \textbf{642}, 436 (2006)
\bibitem{James14} J.M. Lattimer, A.W. Steiner, Eur. Phys. J. A \textbf{50}, 40 (2014)
\bibitem{Richter23} J. Richter, B.A. Li, Phys. Rev. C \textbf{108}, 055803 (2023)
\bibitem{LCK} B.A. Li, L.W. Chen, C.M. Ko, Phys. Rep. \textbf{464}, 113 (2008)
\bibitem{EPJA-review} B.A. Li, P. G. Krastev, D.H. Wen, N.B. Zhang, Eur. Phys. J. A \textbf{55}, 117 (2019)
\bibitem{Essick24} R. Essick, Astrophys. J. Lett. \textbf{973}, no.2, L50 (2024)
\bibitem{Baym18} G. Baym, T. Hatsuda, T. Kojo, P.D. Powell, Y. Song, T. Takatsuka, Rep. Prog. Phys. \textbf{81}, 056902 (2018)
\bibitem{Hempel16} M. Hempel, O. Heinimann, A. Yudin, I. Iosilevskiy, M. Liebend\"{o}rfer, F.K. Thielemann, Phys. Rev. D \textbf{94}, 103001 (2016)
\bibitem{Carlomagno24} J.P. Carlomagno, G.A. Contrera, A.G. Grunfeld, D. Blaschke, Phys. Rev. D \textbf{109}, 043050 (2024)}
\end{thebibliography}
%

\end{document}